\documentclass[%
 preprint,
 amsmath,amssymb,
 aps, physrev,
]{revtex4-2}

\usepackage{graphicx}
\usepackage{dcolumn}
\usepackage{bm}
\usepackage[utf8]{inputenc}
\usepackage[top = 1in, bottom = 1in, right= 1in, left = 1in]{geometry}
\usepackage{physics}
\usepackage{amsmath}
\usepackage{amssymb}
\usepackage{natbib}
\begin{document}

\title{\textbf{Optimal propagation distance for maximal biphoton entanglement through the weakly turbulent atmosphere} 
}%

\author{Luchang Niu}
\affiliation{Department of Physics and Astronomy, University of Rochester, Rochester, New York 14627, USA}
\email{Contact author: lniu5@u.rochester.edu}

\author{Yang Xu}
\affiliation{Department of Physics and Astronomy, University of Rochester, Rochester, New York 14627, USA}

\author{Saleem Iqbal}
\affiliation{The Institute of Optics, University of Rochester, Rochester, New York 14627, USA}

\author{Robert W. Boyd}
\affiliation{The Institute of Optics, University of Rochester, Rochester, New York 14627, USA}
\affiliation{Department of Physics and Astronomy, University of Rochester, Rochester, New York 14627, USA}
\affiliation{Department of Physics, University of Ottawa, Ottawa, Ontario K1N 6N5, Canada}

\date{\today}

\begin{abstract}
Understanding the influence of atmospheric turbulence on the propagation of entangled biphoton states is essential for free-space quantum communication protocols. Using the extended Huygens-Fresnel principle and the Kolmogorov turbulence model, we derive an analytical expression for the combined density operator of the signal and idler fields generated via SPDC, following propagation through separate turbulent channels. By expressing this density operator in the continuous position basis, we show how the spatial correlations between signal and idler persist through turbulence despite the loss of state purity, as they transition from being quantum to classical in nature. We further identify a finite range of propagation distances over which the angle-OAM entanglement is maximized, which provides valuable insights for designing free-space quantum communication links operating over several kilometers through the turbulent atmosphere.
\end{abstract}

\maketitle

\section{Introduction}
Optical beams can encode and transmit information in various degrees of freedom, including polarization, frequency, and spatial modes. Due to its high information capacity, free-space optical (FSO) communication has emerged as a practical wireless technology that transmits data through the atmosphere, offering high bandwidth without the need for physical transmission cables. In recent years, the use of quantum information approaches for FSO communication protocols has received broad interest \cite{tsujino2011quantum,jiang2007fast,yang2021towards,martinez2018high,cao2020long,yang2024high,chen2021twin,cozzolino2019orbital,tan2024real}. The implementation of these protocols exploits different degrees of freedom, such as the polarization \cite{tan2024real,dahal2021polarization} and orbital angular momentum (OAM) \cite{cozzolino2019orbital,mafu2013higher, xu2024stimulated} of light, to enhance security, resilience, and efficiency in data transmission \cite{tsujino2011quantum,jiang2007fast,lo2014secure}. However, the quantum states employed in these protocols are particularly vulnerable to atmospheric turbulence, which induces decoherence and modal crosstalk, thereby degrading the fidelity of the quantum states and the overall system performance. Accurate modeling of atmospheric turbulence and a rigorous characterization of its effects on nonclassical states of light are thus essential for developing high-fidelity, free-space quantum communication protocols. In recent years, extensive studies have investigated the effects of turbulence or other types of random media on quantum beams, particularly focusing on orbital angular momentum (OAM) mode distortion and crosstalk \cite{tyler2009influence,yan2016decoherence}, as well as entanglement degradation \cite{jha2010effects,roux2015entanglement,ibrahim2014parameter}. 

However, previous studies of the effects of turbulence on two-photon entanglement have typically simplified their models by truncating the otherwise continuous spatial degrees of freedom to a small discrete basis. This level of simplification fails to capture the statistical behavior of spatial field distributions and correlations. Furthermore, these studies often overly simplify the continuous evolution of the spatial structure of the field as it propagates. For example, it has been shown that two-photon entanglement can migrate between amplitude and phase degrees of freedom during free-space propagation even in the absence of turbulence \cite{chan2007transverse,reichert2017quality}. Moreover, propagation can also redistribute entanglement across different bases \cite{bhattacharjee2022propagation,just2013transverse}. It is therefore essential to incorporate entanglement migration and redistribution in studies of two-photon state propagation through turbulence, and to identify ways to exploit them to mitigate the detrimental effects of turbulence in FSO communication. To address these challenges, we develop a modeling framework, based on the extended Huygens-Fresnel principle and the Kolmogorov model for turbulence, to study the evolution of the spatial structure and entanglement of the entire two-photon field. We consider SPDC-generated entangled photon pairs propagating through atmospheric turbulence. Rather than truncating to a two-qubit approximation, we retain the full spatial structure of the two-photon field by employing a continuous-variable and infinite-dimensional density matrix in the position basis. This framework enables analysis of the full spatial state propagation and the role of entanglement redistribution in turbulent channels, offering deeper insights into the maintenance of quantum communication fidelity in realistic environments.

This paper is organized as follows. In Section \ref{sec:method}, we present the extended Huygens-Fresnel (eHF) principle and apply it to the propagation of a two-photon field produced by SPDC and collimated with a thin lens. In Section \ref{sec:main_section}, we characterize the effects of turbulence on the two-photon field, and identify a finite range of propagation distances over which the angle–OAM entanglement is maximized. A summary and conclusions are then provided in Section \ref{sec:discussion}.

\section{Method}
\label{sec:method}
\subsection{Extended Huygens–Fresnel (eHF) principle }
\label{subsec:eHF}
The Huygens–Fresnel (HF) principle states that every point on a wavefront acts as a secondary source of spherical wavelets. The new wavefront is the superposition of these wavelets. The extended Huygens–Fresnel (eHF) principle  \cite{wang1979optical,andrews2005laser,wang2015propagation} is a generalized version of the conventional Huygens–Fresnel principle. It incorporates a statistical model of atmospheric turbulence to account for the random distortion of the wavefront. It is especially suitable for analyzing beam propagation through weakly turbulent media. Several studies have demonstrated that the eHF provides a reliable and effective model for light propagation through turbulence \cite{andrews2005laser,bochove2016approach,ponomarenko2022classical}. 

The eHF principle states that the optical field $U(\boldsymbol{\rho}, z)$ after propagating a distance $z$ through a turbulent medium is given by \cite{wang2015propagation}:
\begin{equation}
U(\boldsymbol{\rho}, z) = -\frac{ik}{2\pi z} \int d^2\boldsymbol{\rho}' \, U(\boldsymbol{\rho}', 0) 
\exp\left(\frac{ik (\boldsymbol{\rho} - \boldsymbol{\rho}')^2}{2z}\right) 
\exp\left[\phi(\boldsymbol{\rho}', \boldsymbol{\rho}, z)\right]
\label{eq:eHF_field}
\end{equation}
where $k=\frac{2\pi}{\lambda}$ and $\phi(\boldsymbol{\rho}', \boldsymbol{\rho}, z)$ is the random part of the complex phase of each spherical wave component due to turbulence. The quantities $\boldsymbol{\rho}'$ and $\boldsymbol{\rho}$ denote the position vectors in the transmitter plane and the receiver plane, respectively, with $U(\boldsymbol{\rho}', 0)$ denoting the light field at the transmitter plane. From equation (\ref{eq:eHF_field}), we can express the cross-spectral density of the field, defined as the average over the realizations of the turbulent channel \cite{yura1972mutual,phehlukwayo2020influence}:
\begin{equation}
\begin{aligned}
W(\boldsymbol{\rho}_1, \boldsymbol{\rho}_2, z) &= \left\langle U^*(\boldsymbol{\rho}_1, z) U(\boldsymbol{\rho}_2, z) \right\rangle_m\\
&= -\frac{ik}{2\pi z} \iint W(\boldsymbol{\rho}_1', \boldsymbol{\rho}_2', 0) 
\exp\left[-\frac{ik (\boldsymbol{\rho}_1 - \boldsymbol{\rho}_1')^2}{2z}\right] \\
&\quad \times \exp\left[\frac{ik (\boldsymbol{\rho}_2 - \boldsymbol{\rho}_2')^2}{2z}\right] \\
&\quad \times \left\langle \exp\left[\phi^*(\boldsymbol{\rho}_1', \boldsymbol{\rho}_1, z) + \phi(\boldsymbol{\rho}_2', \boldsymbol{\rho}_2, z)\right] \right\rangle_m \, d^2\boldsymbol{\rho}_1' d^2\boldsymbol{\rho}_2
\label{eq:eHF_spectrum}
\end{aligned}
\end{equation}
where the subscript ``m" indicates the ensemble average over the turbulent medium. For homogeneous and isotropic turbulence, we apply the Kolmogorov model and use a quadratic approximation \cite{wang1979optical} to obtain an analytical expression of the turbulence function \cite{wang2015propagation}:
\begin{equation}
\left\langle \exp \left[ \phi^*(\boldsymbol{\rho}_1', \boldsymbol{\rho}_1, z) + \phi(\boldsymbol{\rho}_2', \boldsymbol{\rho}_2, z) \right] \right\rangle_m 
= \exp \left[ -\frac{1}{\rho_0^2} \left( \boldsymbol{\rho}_d'^2 + \boldsymbol{\rho}_d' \cdot \boldsymbol{\rho}_d + \boldsymbol{\rho}_d^2 \right) \right]
\label{eq:turbulence function}
\end{equation}
where $\boldsymbol{\rho}_d=\boldsymbol{\rho}_1-\boldsymbol{\rho}_2$, $\boldsymbol{\rho}_d'=\boldsymbol{\rho}_1'-\boldsymbol{\rho}_2'$, $\rho_0 = \left( 0.546 C_n^2 k^2 z \right)^{-3/5}$ is the coherence length of a spherical wave propagating in turbulence, and $C_n^2$ is the refractive index structure constant \cite{wang2015propagation,yura1972mutual}. 

\subsection{Propagation of the Two-photon Field Through Turbulence}
\label{subsec:propagation}
We have established the eHF principle as an effective method to describe field propagation through turbulence. Now, we apply this method to characterize the propagation of a two-photon field generated by type-I SPDC through turbulence. For a degenerate SPDC process where the collinear phase-matching condition is satisfied, the two-photon field at $z=0$ in the position basis is given by \cite{schneeloch2016introduction,walborn2010spatial,edgar2012imaging}:
\begin{equation}
\psi(\boldsymbol{\rho}_s, \boldsymbol{\rho}_i; 0) = A \exp \left[ -\frac{(\boldsymbol{\rho}_s + \boldsymbol{\rho}_i)^2}{4w_0^2} \right]
\exp \left[ -\frac{(\boldsymbol{\rho}_s - \boldsymbol{\rho}_i)^2}{4\sigma_0^2} \right]
\label{eq:spdc field}
\end{equation}
where we have used the double Gaussian approximation, $w_0$ is the beam waist of the Gaussian pump, $\boldsymbol{\rho}_s$ and $\boldsymbol{\rho}_i$ are the position vectors for the signal and idler fields, $\sigma_0=\sqrt{\frac{0.455 L \lambda_p}{2\pi}}$ is the transverse spatial correlation length, $L$ is the thickness of the nonlinear crystal, and $A=\frac{1}{\pi w_0 \sigma_0}$ is a normalization factor. For simplicity, we will use a formal variable $A$ to represent all normalization factors regardless of their exact values for different expressions from now on.

\begin{figure}
    \centering
    \includegraphics[width=1\linewidth]{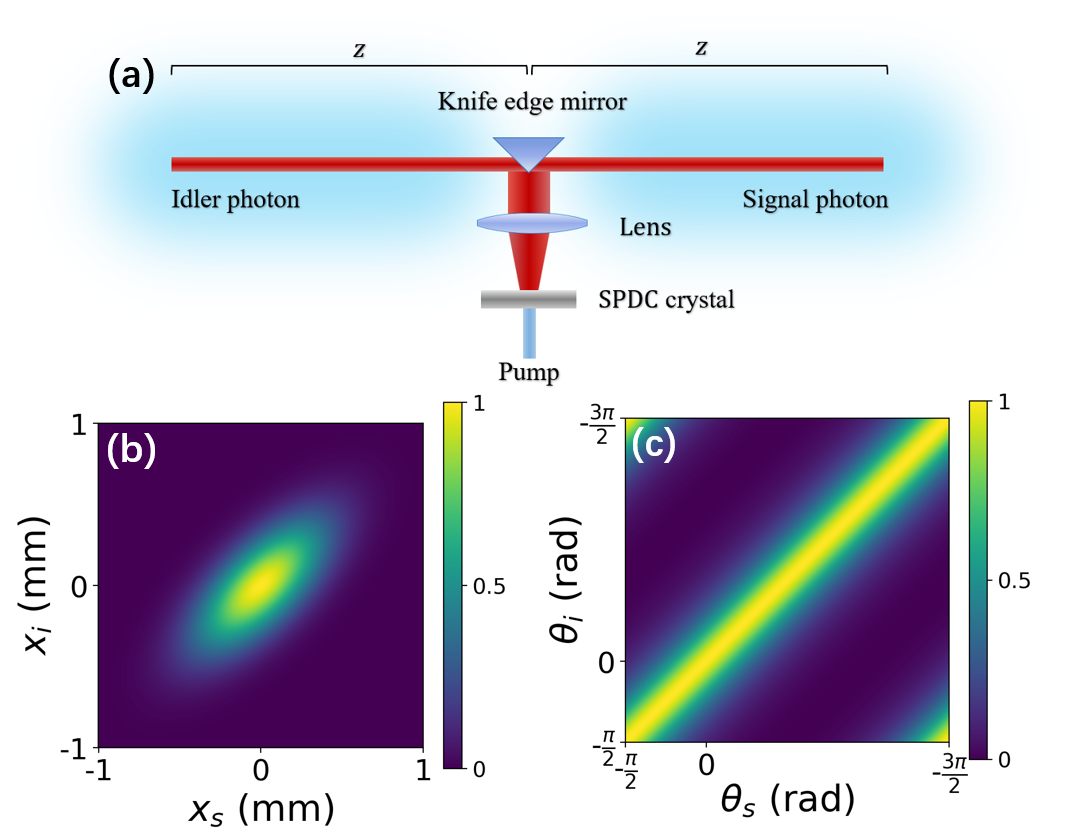}
    \caption{Propagation of a two-photon field through turbulence. (a) Schematic illustrations of the SPDC process and the field propagation. (b) Joint two-photon position probability distribution function $P(x_s,x_i;z)$ at the output face of the crystal before collimation. (c) Joint two-photon angle probability distribution function $P(\theta_s,\theta_i;z)$ in the same plane, where $\theta$ is the azimuthal angle. The tight correlations in both the position and angle bases can be clearly seen.}
    \label{fig:p_distribution}
\end{figure}

The SPDC process and the propagation of the two-photon field are illustrated in figure \ref{fig:p_distribution}. After generation, the signal and idler photons are separated and then propagate along different paths. The second-order cross-spectral density function of the two-photon field after propagation through a turbulent medium of a distance $z$ is given by \cite{wang2015propagation,wang1983receiver}: 
\begin{equation} \begin{aligned} 
W_2&(\boldsymbol{\rho}_{s1},\boldsymbol{\rho}_{i1}, \boldsymbol{\rho}_{s2},\boldsymbol{\rho}_{i2}, z)\\
=&(-\frac{ik}{2\pi z})^2 \iiiint W_2(\boldsymbol{\rho}_{s1}',\boldsymbol{\rho}_{i1}', \boldsymbol{\rho}_{s2}',\boldsymbol{\rho}_{i2}', 0) \\ 
&\times \exp\left[\frac{ik (\boldsymbol{\rho}_{s2} - \boldsymbol{\rho}_{s2}')^2}{2z}+\frac{ik (\boldsymbol{\rho}_{i2} - \boldsymbol{\rho}_{i2}')^2}{2z}\right] \\ 
&\times \left \langle \exp\left[ \phi^*(\boldsymbol{\rho}_{s1}', \boldsymbol{\rho}_{s1}, z) + \phi(\boldsymbol{\rho}_{s2}', \boldsymbol{\rho}_{s2}, z) + \phi^*(\boldsymbol{\rho}_{i1}', \boldsymbol{\rho}_{i1}, z) + \phi(\boldsymbol{\rho}_{i2}', \boldsymbol{\rho}_{i2}, z) \right] \right\rangle_m
\label{eq:eHF_spdc} \end{aligned} \end{equation} 
Since the signal and idler photons propagate through statistically independent turbulent channels, the turbulence-induced random phases of the two photons are spatially uncorrelated. Therefore, the random phase factor in equation (\ref{eq:eHF_spdc}) can be written as:
\begin{equation} \begin{aligned} 
&\left\langle \exp\left[ \phi^*(\boldsymbol{\rho}_{s1}', \boldsymbol{\rho}_{s1}, z) + \phi(\boldsymbol{\rho}_{s2}', \boldsymbol{\rho}_{s2}, z) + \phi^*(\boldsymbol{\rho}_{i1}', \boldsymbol{\rho}_{i1}, z) + \phi(\boldsymbol{\rho}_{i2}', \boldsymbol{\rho}_{i2}, z) \right] \right\rangle_m \\ 
& \approx \left\langle \exp \left[ \phi^*(\boldsymbol{\rho}_{s1}', \boldsymbol{\rho}_{s1}, z) + \phi(\boldsymbol{\rho}_{s2}', \boldsymbol{\rho}_{s2}, z) \right] \right\rangle_m \cdot \left\langle \exp \left[ \phi^*(\boldsymbol{\rho}_{i1}', \boldsymbol{\rho}_{i1}, z) + \phi(\boldsymbol{\rho}_{i2}', \boldsymbol{\rho}_{i2}, z) \right] \right\rangle_m \\ 
&=\exp \left[ -\frac{\left( \boldsymbol{\rho}_{ds}'^2 + \boldsymbol{\rho}_{ds}' \cdot \boldsymbol{\rho}_{ds} + \boldsymbol{\rho}_{ds}^2 \right)+\left( \boldsymbol{\rho}_{di}'^2 + \boldsymbol{\rho}_{di}' \cdot \boldsymbol{\rho}_{di} + \boldsymbol{\rho}_{di}^2 \right)}{\rho_0^2} \right] 
\end{aligned} \end{equation} 
where we have defined $\boldsymbol{\rho}_{ds}=\boldsymbol{\rho}_{s1}-\boldsymbol{\rho}_{s2}$, $\boldsymbol{\rho}_{ds}'=\boldsymbol{\rho}_{s1}'-\boldsymbol{\rho}_{s2}'$, $\boldsymbol{\rho}_{di}=\boldsymbol{\rho}_{i1}-\boldsymbol{\rho}_{i2}$, $\boldsymbol{\rho}_{di}'=\boldsymbol{\rho}_{i1}'-\boldsymbol{\rho}_{i2}'$, and made use of equation (\ref{eq:turbulence function}). The second-order cross-spectral density function at $z=0$ is given by $W_2(\boldsymbol{\rho}'_{s1},\boldsymbol{\rho}'_{i1}, \boldsymbol{\rho}'_{s2},\boldsymbol{\rho'}_{i2}, 0)=\psi(\boldsymbol{\rho}'_{s2}, \boldsymbol{\rho}'_{i2}; 0) \psi^*(\boldsymbol{\rho}'_{s1}, \boldsymbol{\rho}'_{i1}; 0)$.

The double Gaussian approximation allows us to separate equation (\ref{eq:spdc field}) into independent $x$ and $y$ components: 
\begin{equation}
\begin{aligned}
\psi(\boldsymbol{\rho}_s, \boldsymbol{\rho}_i; 0) 
&= A \exp \left[ -\frac{(\boldsymbol{\rho}_s + \boldsymbol{\rho}_i)^2}{4w_0^2} \right] 
\exp \left[ -\frac{(\boldsymbol{\rho}_s - \boldsymbol{\rho}_i)^2}{4\sigma_0^2} \right]\\
&=A \exp \left[ -\frac{(x_s + x_i)^2+(y_s + y_i)^2}{4w_0^2} \right] 
\exp \left[ -\frac{(x_s - x_i)^2+(y_s - y_i)^2}{4\sigma_0^2} \right]\\
&= \psi(x_s, x_i; 0)\psi(y_s, y_i; 0).
\label{eq:decompose}
\end{aligned}
\end{equation}
Because $W_2(\boldsymbol{\rho}_{s1},\boldsymbol{\rho}_{i1},\boldsymbol{\rho}_{s2},\boldsymbol{\rho}_{i2}, z) =W_2(x_{s1},x_{i1}, x_{s2},x_{i2}, z) W_2(y_{s1},y_{i1}, y_{s2},y_{i2}, z)$, a similar decomposition can be applied to equation (\ref{eq:eHF_spdc}):
\begin{equation}
\begin{aligned}
W_2(x_{s1},x_{i1}, x_{s2},x_{i2}, z) 
&\propto \iiiint W_2(x_{s1},x_{i1}, x_{s2},x_{i2}, 0) \\
&\quad \times\exp\left[\frac{ik (x_{s2} - x_{s2}')^2}{2z}+\frac{ik (x_{i2} - x_{i2}')^2}{2z}\right] \\
&\quad \times \exp\left[-\frac{ik (x_{s1} - x_{s1}')^2}{2z}-\frac{ik (x_{i1} - x_{i1}')^2}{2z}\right] \\
&\quad \times \exp \left[ -\frac{\left( x_{ds}'^2 + x_{ds}' \cdot x_{ds} + x_{ds}^2 \right)+\left( x_{di}'^2 + x_{di}' \cdot x_{di} + x_{di}^2 \right)}{\rho_0^2}  \right]\\
&\quad d^2 x_{s1}' d^2 x_{s2}' d^2 x_{i1}' d^2 x_{i2}'
\label{eq:eHF_x}
\end{aligned}
\end{equation}
where the normalization factor is omitted for brevity. The $x$-component $W(x_{s1},x_{i1}, x_{s2},x_{i2}, z)$ has the same form as $W(y_{s1},y_{i1}, y_{s2},y_{i2}, z)$ due to symmetry, thus $W(\boldsymbol{\rho}_{s1},\boldsymbol{\rho}_{i1}, \boldsymbol{\rho}_{s2},\boldsymbol{\rho}_{i2}, z) $ can be obtained by solving equation (\ref{eq:eHF_x}).

To simplify the integral, the following variable substitutions are introduced: 
\begin{equation}
\begin{aligned}
\boldsymbol{\rho}_{2+}=\frac{\boldsymbol{\rho}_{2s}+\boldsymbol{\rho}_{2i}}{\sqrt{2}},\quad 
\boldsymbol{\rho}_{2-}=\frac{\boldsymbol{\rho}_{2s}-\boldsymbol{\rho}_{2i}}{\sqrt{2}},\quad  
\boldsymbol{\rho}_{1+}=\frac{\boldsymbol{\rho}_{1s}+\boldsymbol{\rho}_{1i}}{\sqrt{2}},\quad 
\boldsymbol{\rho}_{1+}=\frac{\boldsymbol{\rho}_{1s}-\boldsymbol{\rho}_{1i}}{\sqrt{2}}
\label{equ:coord_trans}
\end{aligned}
\end{equation}
After evaluating the integral of equation (\ref{eq:eHF_x}), the second-order cross-spectral density function at a propagation distance $z$ is obtained:
\begin{equation}
\begin{aligned}
W_2(\boldsymbol{\rho}_{2+}, \boldsymbol{\rho}_{1+}, \boldsymbol{\rho}_{2-}, \boldsymbol{\rho}_{1-},z) 
&=|A|^2 \exp \left[ -\frac{\boldsymbol{\rho}_{2+}^2 + \boldsymbol{\rho}_{1+}^2}{2w^2(z)} \right]
 \exp \left[ -\frac{\boldsymbol{\rho}_{2-}^2 + \boldsymbol{\rho}_{1-}^2}{2\sigma^2(z)} \right] \\
&\quad \times \exp \left[ i k \frac{\boldsymbol{\rho}_{2+}^2 - \boldsymbol{\rho}_{1+}^2}{2R_+^2(z)} \right]
 \exp \left[ i k \frac{\boldsymbol{\rho}_{2-}^2 - \boldsymbol{\rho}_{1-}^2}{2R_-^2(z)} \right] \\
&\quad \times \exp \left[ -\frac{(\boldsymbol{\rho}_{2+} - \boldsymbol{\rho}_{1+})^2}{C_+^2(z)} \right]
 \exp \left[ -\frac{(\boldsymbol{\rho}_{2-} - \boldsymbol{\rho}_{1-})^2}{C_-^2(z)} \right]
\label{eq:W_tur}
\end{aligned}
\end{equation}
where 
\begin{equation}
\begin{aligned}
&w^2(z)=w_0^2 + \frac{z^2}{k^2}( \frac{1}{w_0^2}+\frac{4}{\rho_0^2} )\\
&\sigma^2(z)=\sigma_0^2 + \frac{z^2}{k^2}( \frac{1}{\sigma_0^2}+\frac{4}{\rho_0^2} )\\
&R_+^2(z)=\frac{k^2 \rho_0^2 w_0^4 + \left(\rho_0^2 + 4 w_0^2\right) z^2}{\left(\rho_0^2 + 6 w_0^2\right) z}\\
&R_-^2(z)=\frac{k^2 \rho_0^2 \sigma_0^4 + \left(\rho_0^2 + 4 \sigma_0^2\right) z^2}{\left(\rho_0^2 + 6 \sigma_0^2\right) z}\\
&C_+^2(z)=\frac{k^2 \rho_0^4 w_0^4 + \rho_0^2 \left(\rho_0^2 + 4 w_0^2\right) z^2}{3 k^2 \rho_0^2 w_0^4 +(\rho_0^2+3 w_0^2 )z^2}\\
&C_-^2(z)=\frac{k^2 \rho_0^4 \sigma_0^4 + \rho_0^2 \left(\rho_0^2 + 4 \sigma_0^2\right) z^2}{3 k^2 \rho_0^2 \sigma_0^4 +(\rho_0^2+3\sigma_0^2 )z^2}
\label{eq:Wtur_para}
\end{aligned}
\end{equation}

For reference, equation (\ref{eq:W_tur}) is compared with the second-order cross-spectral density function after propagating a distance $z$ in the absence of turbulence, which is given by \cite{schneeloch2016introduction,reichert2017quality}: 
\begin{equation}
\begin{aligned}
W_{2f}(\boldsymbol{\rho}_{2+}, \boldsymbol{\rho}_{1+}, \boldsymbol{\rho}_{2-}, \boldsymbol{\rho}_{1-},z)
&=|A|^2\exp \left[ -\frac{\boldsymbol{\rho}_{2+}^2+\boldsymbol{\rho}_{1+}^2}{2w_f^2(z)} \right]
\exp \left[ -\frac{\boldsymbol{\rho}_{2-}^2+\boldsymbol{\rho}_{1-}^2}{2\sigma_f^2(z)} \right]\\
&\quad \times \exp \left[ i k \frac{\boldsymbol{\rho}_{2+}^2 - \boldsymbol{\rho}_{1+}^2}{2R_{f+}^2(z)} \right]
\exp \left[ i k \frac{\boldsymbol{\rho}_{2-}^2 - \boldsymbol{\rho}_{1-}^2}{2R_{f-}^2(z)} \right]
\label{eq:W_free}
\end{aligned}
\end{equation}
In equation (\ref{eq:W_tur}), the first two terms describe the transverse intensity distribution and spatial correlations of the field. The two middle terms represent the phase accumulated during propagation. The last two terms represent the cross-coupling that results from the scattering effect of turbulence. Equation (\ref{eq:W_tur}) reduces to equation (\ref{eq:W_free}) under the limit $C_n^2\rightarrow0$ or $\boldsymbol{\rho}_0\rightarrow+\infty$, so that $C_+^2(z)\rightarrow+\infty$, $C_-^2(z)\rightarrow+\infty$, $R_{-}^2(z)\rightarrow R_{f-}^2(z)$, $R_{+}^2(z)\rightarrow R_{f+}^2(z)$, $w^2(z)\rightarrow w_f^2(z)=w_0^2 (1 + \frac{z^2}{k^2 w_0^4})$, and $\sigma^2(z)\rightarrow \sigma_f^2(z)=\sigma_0^2  (1 + \frac{z^2}{k^2 \sigma_0^4})$. It is worth noting that the second-order cross-spectral density function in equation (\ref{eq:W_tur}) is mathematically equivalent to the density matrix elements of the biphoton state in the position representation. The density matrix becomes ``more diagonal" as the turbulence strength $C_n^2$ or the propagation distance $z$ increases. As the coupling parameters $C_+(z)$ and $C_-(z)$ decrease, the off-diagonal terms ($\boldsymbol{\rho}_{2+} \ne \boldsymbol{\rho}_{1+}$ or $\boldsymbol{\rho}_{2-} \ne \boldsymbol{\rho}_{1-}$) in the density matrix decay more rapidly. This phenomenon is commonly known as decoherence, and indicates the gradual evolution of the initial two-photon state from a pure state into a mixed state as the field propagates through turbulence.

The joint probability distribution for the transverse positions of the signal and idler photons in each pair can be directly measured through coincidence detection, and can be obtained by setting $\boldsymbol{\rho}_{2+}=\boldsymbol{\rho}_{1+}$ and $\boldsymbol{\rho}_{2-}=\boldsymbol{\rho}_{1-}$ in the second-order cross-spectral density:
\begin{equation}
\begin{aligned}
P(\boldsymbol{\rho}_{+},\boldsymbol{\rho}_{-},z)
&= |A|^2\exp \left[ -\frac{\boldsymbol{\rho}_{+}^2 }{\sigma_+^2(z)} \right]
 \exp \left[ -\frac{\boldsymbol{\rho}_{-}^2}{\sigma_-^2(z)} \right]\\
&=|A|^2\exp \left[ -\frac{(\boldsymbol{\rho}_{s}+\boldsymbol{\rho}_{i})^2 }{2w^2(z)} \right]
\exp \left[ -\frac{(\boldsymbol{\rho}_{s}-\boldsymbol{\rho}_{i})^2}{2\sigma^2(z)} \right].
\label{eq:Px}
\end{aligned}
\end{equation}
Using the coordinate transformation $|\boldsymbol{\rho}_s + \boldsymbol{\rho}_i|^2 = r_s^2 + r_i^2 + 2r_s r_i \cos(\theta_s - \theta_i)$ and $|\boldsymbol{\rho}_s - \boldsymbol{\rho}_i|^2 = r_s^2 + r_i^2 - 2r_s r_i \cos(\theta_s - \theta_i) $, the joint probability distribution can be expressed in polar coordinates as:
\begin{equation}
\begin{aligned}
P(r_s, \theta_s, r_i, \theta_i; z) &= |A|^2 \exp\left[ -\frac{r_s^2 + r_i^2 + 2r_s r_i \cos(\theta_s - \theta_i)}{2w(z)^2} \right] \\
& \quad \times \exp\left[ -\frac{r_s^2 + r_i^2 - 2r_s r_i \cos(\theta_s - \theta_i)}{2\sigma(z)^2} \right]
\label{equ:coordinate}
\end{aligned}
\end{equation}
so that the two-photon angle distribution can be obtained by tracing out the radial coordinate:
\begin{equation}
P(\theta_s, \theta_i; z) = \iint r_s r_i P(r_s, \theta_s, r_i, \theta_i; z) \, dr_s \, dr_i
\label{equ:P_angle}
\end{equation}

\subsection{Angle-OAM Entanglement in Turbulence}
\label{subsec:entanglement}
To investigate the evolution of the angle-OAM entanglement during propagation in turbulence, we employ the EPR criterion in the angle-OAM bases \cite{leach2010quantum,schneeloch2018quantifying,bhattacharjee2022propagation}:
\begin{equation}
\Delta(\theta_s \mid \theta_i; z) \Delta(\ell_s \mid \ell_i; z) \geq 0.5 \hbar \left[ 1 - 2\pi P(\theta_0 \mid \theta_i; z) \right].
\label{equ:epr_angle}
\end{equation}
The system is entangled when the inequality is violated. The degree of entanglement is quantified by the extent to which the left-hand side falls below the right-hand side. The conditional angle uncertainty $\Delta (\theta_s|\theta_i;z)$ can be obtained from the two-photon angle probability distribution in equation (\ref{equ:P_angle}) by setting $\theta_i=0$. Similarly, for the conditional OAM distribution, we choose $\ell_i = 0$. 

The conditional OAM distribution can be obtained from the density matrix following propagation through the two turbulent channels by changing from the transverse position basis to the Laguerre-Gauss mode basis. The joint probability distribution for the radial mode indices $p_s$, $p_i$ and the azimuthal mode indices $l_s$, $l_i$ of the signal and idler photons can then be expressed as:
\begin{equation}
\begin{aligned}
P(l_s,p_s,l_i,p_i;z) =& \iiiint d\boldsymbol{\rho}_{1s}d\boldsymbol{\rho}_{1i}d\boldsymbol{\rho}_{2s}d\boldsymbol{\rho}_{2i}\ W_2\left(\boldsymbol{\rho}_{1s},\boldsymbol{\rho}_{1i},\boldsymbol{\rho}_{2s},\boldsymbol{\rho}_{2i};z\right)\\
&\cdot\psi_{2,LG}\left(\boldsymbol{\rho}_{1s},\boldsymbol{\rho}_{1i}\right)\cdot\psi^*_{2,LG}\left(\boldsymbol{\rho}_{2s},\boldsymbol{\rho}_{2i}\right)
\end{aligned}
\label{eq:overlap}
\end{equation}
where $W_2\left(\boldsymbol{\rho}_{1s},\boldsymbol{\rho}_{1i},\boldsymbol{\rho}_{2s},\boldsymbol{\rho}_{2i};z\right)$ is given by equations (\ref{equ:coord_trans}, \ref{eq:W_tur}), and
\begin{equation}
\begin{split}
    \psi_{2,LG}\left(\boldsymbol{\rho}_{s},\boldsymbol{\rho}_{i}\right) = LG_{l_s,p_s}\left(\boldsymbol{\rho}_s\right)\cdot LG_{l_i,p_i}\left(\boldsymbol{\rho}_i\right)
\end{split}
\end{equation}
is the two-photon wavefunction for a separable state of a pair of signal and idler photons in two arbitrary Laguerre-Gauss modes. Subsequently, the two-photon joint OAM distribution can be obtained by tracing over the radial mode indices:
\begin{equation}
    P(l_s,l_i;z) = \sum_{p_s=0}^\infty\sum_{p_i=0}^\infty P(l_s,p_s,l_i,p_i;z)
\label{eq:sum_p}
\end{equation}
and then the conditional OAM distribution is obtained by setting $l_i=0$. 

The following identity is useful in simplifying equation (\ref{eq:sum_p}):
\begin{equation}
    \sum_{p=0}^{\infty}\frac{2p!}{\pi\left(p + \left|l\right|\right)!}L_p^{\left|l\right|}\left(\frac{2\rho_1^2}{w^2}\right)L_p^{\left|l\right|}\left(\frac{2\rho_2^2}{w^2}\right) = \frac{w^2}{2\pi\rho_m}\left(\frac{w^2}{2\rho_1\rho_2}\right)^{\left|l\right|}e^{\frac{\rho_1^2 + \rho_2^2}{w^2}}\delta\left(\rho_1-\rho_2\right)
\label{eq:identity}
\end{equation}
where $L_p^{|l|}(x)$ is a generalized Laguerre polynomial, $w$ is the beam waist, and the index $m = 2$ when the delta function is used to evaluate an integral over $\rho_1$ and vice-versa.

After insertion of equation (\ref{eq:overlap}) into equation (\ref{eq:sum_p}), the identity in equation (\ref{eq:identity}) allows for the evaluation of both sums and two of the integrals once the ordering of the summations and integrations are interchanged. After setting $l_i=0$ and rewriting the integrand in polar coordinates, the conditional OAM distribution is then
\begin{equation}
\begin{aligned}
P\left(l_s|l_i=0;z\right) = &\iiiint_0^{2\pi} d\theta_{1s}d\theta_{2s}d\theta_{1i}d\theta_{2i}\iint_0^{+\infty} d\rho_sd\rho_i\frac{\rho_s\rho_i}{4\pi^2}\ \\
&\cdot W_2\left(\rho_s, \rho_i, \theta_{1s}, \theta_{2s}, \theta_{1i}, \theta_{2i}; z\right)e^{-il_s\left(\theta_{1s} - \theta_{2s}\right)}
\end{aligned}
\label{eq:Pls}
\end{equation}
where $\rho_{1s} = \rho_{2s} \equiv \rho_s$ and $\rho_{1i} = \rho_{2i} \equiv \rho_i$ in the argument of the second-order cross-spectral density function. The conditional OAM uncertainty can then be obtained subsequently.

\subsection{Collimation with a thin lens}
\label{sec:IID}
In the derivation above, the field was propagated through the turbulence without collimating the rapidly diverging down-conversion field that originates from the crystal. This could lead to a broad intensity distribution and consequently requires a large detector aperture. Therefore, a more realistic case is considered where the two-photon field is collimated by an ideal thin lens of focal length $f$. The input plane is taken as the Fourier plane of the crystal exit face:
\begin{equation}
\begin{aligned}
\psi_{in}(\boldsymbol{\rho}_s^{in}, \boldsymbol{\rho}_i^{in}; 0) &\propto \int \psi(\boldsymbol{\rho}_s, \boldsymbol{\rho}_i; 0) \exp\left[-i2\pi\frac{(\boldsymbol{\rho}_s\boldsymbol{\rho}_s^{in})+(\boldsymbol{\rho}_i\boldsymbol{\rho}_i^{in})}{\lambda f}\right]d\boldsymbol{\rho}_id\boldsymbol{\rho}_s\\
&\propto \exp{\left[-\frac{\pi^2w_0^2}{\lambda^2f^2}(\boldsymbol{\rho}_s^{in}+\boldsymbol{\rho}_i^{in})^2\right]}\exp{\left[-\frac{\pi^2\sigma_0^2}{\lambda^2f^2}(\boldsymbol{\rho}_s^{in}-\boldsymbol{\rho}_i^{in})^2\right]}\\
&=\exp \left[ -\frac{(\boldsymbol{\rho}_s^{in} + \boldsymbol{\rho}_i^{in})^2}{4w_c(0)^2} \right]\exp \left[ -\frac{(\boldsymbol{\rho}_s^{in} - \boldsymbol{\rho}_i^{in})^2}{4\sigma_c(0)^2} \right]
\label{eq:collimation}
\end{aligned}
\end{equation}
where $\psi(\boldsymbol{\rho}_s, \boldsymbol{\rho}_i; 0)=A \exp \left[ -\frac{(\boldsymbol{\rho}_s + \boldsymbol{\rho}_i)^2}{4w_0^2} \right]
\exp \left[ -\frac{(\boldsymbol{\rho}_s - \boldsymbol{\rho}_i)^2}{4\sigma_0^2} \right]$ as given in equation (\ref{eq:eHF_field}) and $\lambda=2\lambda_p$ denotes the common wavelength of the signal and idler fields. Note that equations (\ref{eq:collimation}) and (\ref{eq:eHF_field}), share the same functional form. 
Therefore, to obtain the two-photon wavefunction with the thin-lens collimation accounted for, replacing $w_0=w(0)$ with $w_c(0)=\frac{\lambda f}{2 \pi w(0)}$ and $\sigma_0=\sigma(0)$ with $\sigma_c(0)=\frac{\lambda f}{2 \pi \sigma(0)}$ is sufficient. Similarly, at a distance $z$ from the input plane, $w(z)$ and $\sigma(z)$ are replaced by $w_c(z)$ and $\sigma_c(z)$, respectively. 

\section{Results}
\label{sec:main_section}
\subsection{Quantum-to-classical transition of spatial correlations in Kolmogorov turbulence}

Purity, defined as $\gamma(z) = \text{tr}(\hat{\rho}^2) $ where $\hat{\rho}$ is the density matrix, quantifies the degree of mixedness in a quantum state, taking the value $\gamma=1$ for a pure state and $\gamma<1$ for a mixed state. Its elements in the position basis are given by equation (\ref{eq:W_tur}), and the elements of $\hat{\rho}^2$ in the position basis are given by:
\begin{equation}
W_{p2}(\boldsymbol{\rho}_{2+}, \boldsymbol{\rho}_{1+}, \boldsymbol{\rho}_{2-}, \boldsymbol{\rho}_{1-}) = 
\int W_2(\boldsymbol{\rho}_{2+}, \boldsymbol{\rho}_{+}', \boldsymbol{\rho}_{2-}, \boldsymbol{\rho}_{-}') 
W_2(\boldsymbol{\rho}_{+}', \boldsymbol{\rho}_{1+}, \boldsymbol{\rho}_{-}', \boldsymbol{\rho}_{1-}) 
\, d\boldsymbol{\rho}_{+}' \, d\boldsymbol{\rho}_{-}'
\end{equation}
The state purity is obtained by summing the diagonal elements of $\rho^2$. In the continuous case, the trace is evaluated by performing the integral:
\begin{equation}
\begin{aligned}
\gamma(z) &= \text{tr}(\hat{\rho}^2) = \int d\boldsymbol{\rho}_+ d\boldsymbol{\rho}_- W_{p2}(\boldsymbol{\rho}_+, \boldsymbol{\rho}_+, \boldsymbol{\rho}_-, \boldsymbol{\rho}_-)\\
&=\frac{C^2_-(z) C^2_+(z)}{\left[C_-^2(z) + 4 \sigma_c^2(z)\right]\left[C_+^2(z) + 4w_c^2( z)\right]}
\end{aligned}
\end{equation}
where $w_c(z)$ and $\sigma_c(z)$ have been defined in terms of $w(z)$ and $\sigma(z)$ in subsection \ref{sec:IID}. The parameters $C_+(z)$, $C_-(z)$, $\sigma^2(z)$ and $w^2(z)$ are given in equation (\ref{eq:Wtur_para}).

The degree of spatial correlation between the signal and idler photons in each pair can be quantified using the normalized cross-correlation function $f_c(z)$:
\begin{equation}
\begin{aligned}
f_c(z)& = \frac{\langle \boldsymbol{\rho}_s \boldsymbol{\rho}_i \rangle}{\sqrt{\langle \boldsymbol{\rho}_s^2 \rangle \langle \boldsymbol{\rho}_i^2 \rangle}}\\
& = \frac{\int d\boldsymbol{\rho}_s  d\boldsymbol{\rho}_i \, \boldsymbol{\rho}_s \boldsymbol{\rho}_i P(\boldsymbol{\rho}_s, \boldsymbol{\rho}_i)}{\left[\int d\boldsymbol{\rho}_s d\boldsymbol{\rho}_i \, \boldsymbol{\rho}_s^2 P(\boldsymbol{\rho}_s, \boldsymbol{\rho}_i)\right]\left[\int d\boldsymbol{\rho}_s  d\boldsymbol{\rho}_i \,\boldsymbol{\rho}_i^2 P(\boldsymbol{\rho}_s, \boldsymbol{\rho}_i)\right]}\\
&=1 - \frac{2 \sigma_c^2(z)}{\sigma_c^2( z) + w_c^2(z)}
\end{aligned}
\end{equation}
where $P(\boldsymbol{\rho}_s, \boldsymbol{\rho}_i)$ is given by equation (\ref{eq:Px}). The spatial correlation function $f_c(z)$ captures the degree of correlation between the signal and the idler regardless of whether that correlation is classical or non-classical in nature. Note that $f_c=1$ indicates perfect correlation, $f_c=-1$ indicates perfect anti-correlation, and $f_c=0$ indicates no correlation.

\begin{figure}[htp]
    \centering
    \includegraphics[width=0.9\linewidth]{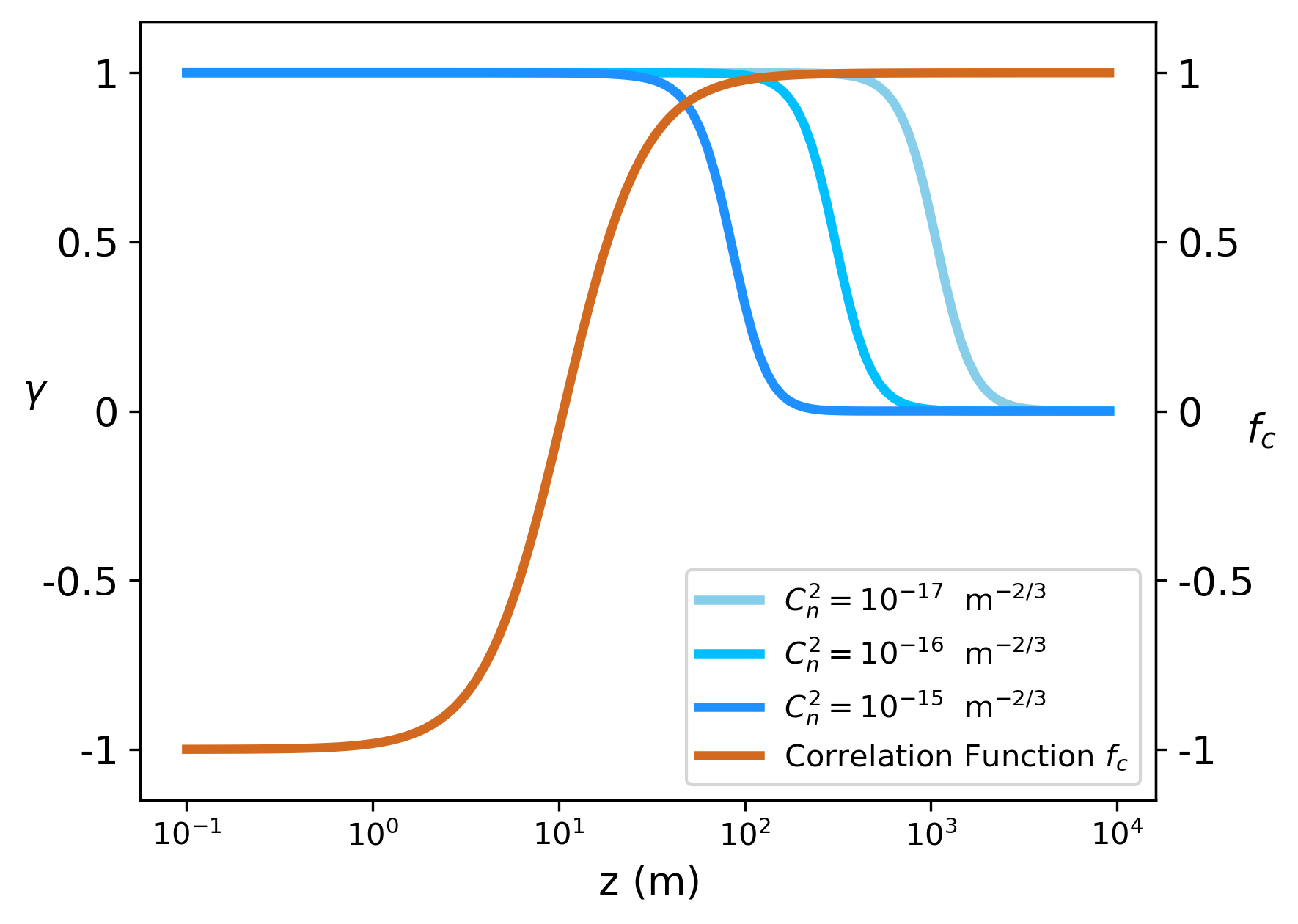}
    \caption{Purity $\gamma(z)$ and spatial correlation function $f_c(z)$ as functions of propagation distance z, under different turbulence strengths. The curves for the correlation function $f_c(z)$ overlap for all turbulence strengths. The parameters are chosen as: crystal length $L=1\,\mathrm{mm}$, focal length $f=50\,\mathrm{cm}$, pump waist $w_0=w(0)=507 \,\mu \mathrm{m}$, and pump wavelength $\lambda_p=355\,\mathrm{nm}$.}
    \label{fig:p_cor}
\end{figure}

Figure \ref{fig:p_cor} shows the purity $\gamma(z)$ and spatial correlation function $f_c(z)$ for different turbulence strengths as a function of the propagation distance $z$. The latter indicates that the signal and idler are initially anti-correlated, and that this spatial anti-correlation gradually transitions to spatial correlation as the field continues to propagate. This effect is not a consequence of the turbulence. Instead, it can be explained by the fact that $w_c(0)\ll\sigma_c(0)$ while $w_c(z)\gg\sigma_c(z)$, which indicates the transition from anti-correlation to correlation. Figure \ref{fig:p_cor} also shows that the $f_c(z)$ curves overlap for all turbulence strengths, whereas the purity decreases more rapidly under stronger turbulence. This suggests that the overall spatial correlation is less affected by the turbulence than the purity. Even when the purity drops to zero, the overall spatial correlation persists.

This persistence of overall spatial correlation can be understood from the statistical properties of the Kolmogorov model: since the turbulence is isotropic and homogeneous, the effects of turbulence on the ensemble of signal photons and the ensemble of idler photons are identical on average. The preservation of the overall correlation together with the complete loss of purity can then be understood as a transition of the spatial correlations from being quantum to classical in nature.

\subsection{Conditional OAM Uncertainty}
\begin{figure}[htp]
    \centering
    \includegraphics[width=1\linewidth]{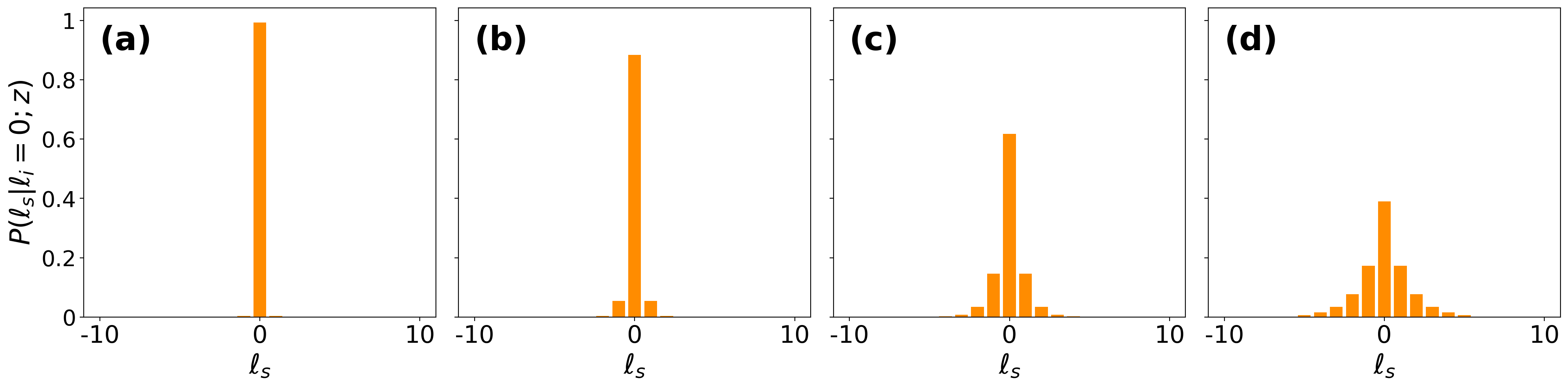}
    \caption{The conditional OAM distribution at different propagation distances $z$. (a) $z=0.5$ km; (b) $z=1$ km; (c) $z=1.5$ km; (d) $z=2$ km. The spectrum broadens with propagation as more OAM modes are coupled together due to the effect of turbulence. The parameters are chosen as: crystal length: $L=1\,\mathrm{mm}$, focal length: $f=50\,\mathrm{cm}$, signal mode waist at \text{z}=0 m: $\sigma_s(0)=507 \,\mu \mathrm{m}$ and pump wavelength: $\lambda_p=355\,\mathrm{nm}$.}
    \label{fig:oam_spectrum_z}
\end{figure}

Below, we use equation (\ref{eq:Pls}) to demonstrate how the conditional OAM spectrum evolves during propagation through turbulence. Figures \ref{fig:oam_spectrum_z}(a)-\ref{fig:oam_spectrum_z}(d) show that the initially narrow conditional OAM distribution broadens with increasing propagation distance through turbulence, indicating the increasingly multimodal nature of the signal field. However, the conserved symmetry of the OAM spectrum ensures that the mean value of the conditional OAM distribution for the signal field remains zero. 

\begin{figure}[htp]
    \centering
    \includegraphics[width=0.9\linewidth]{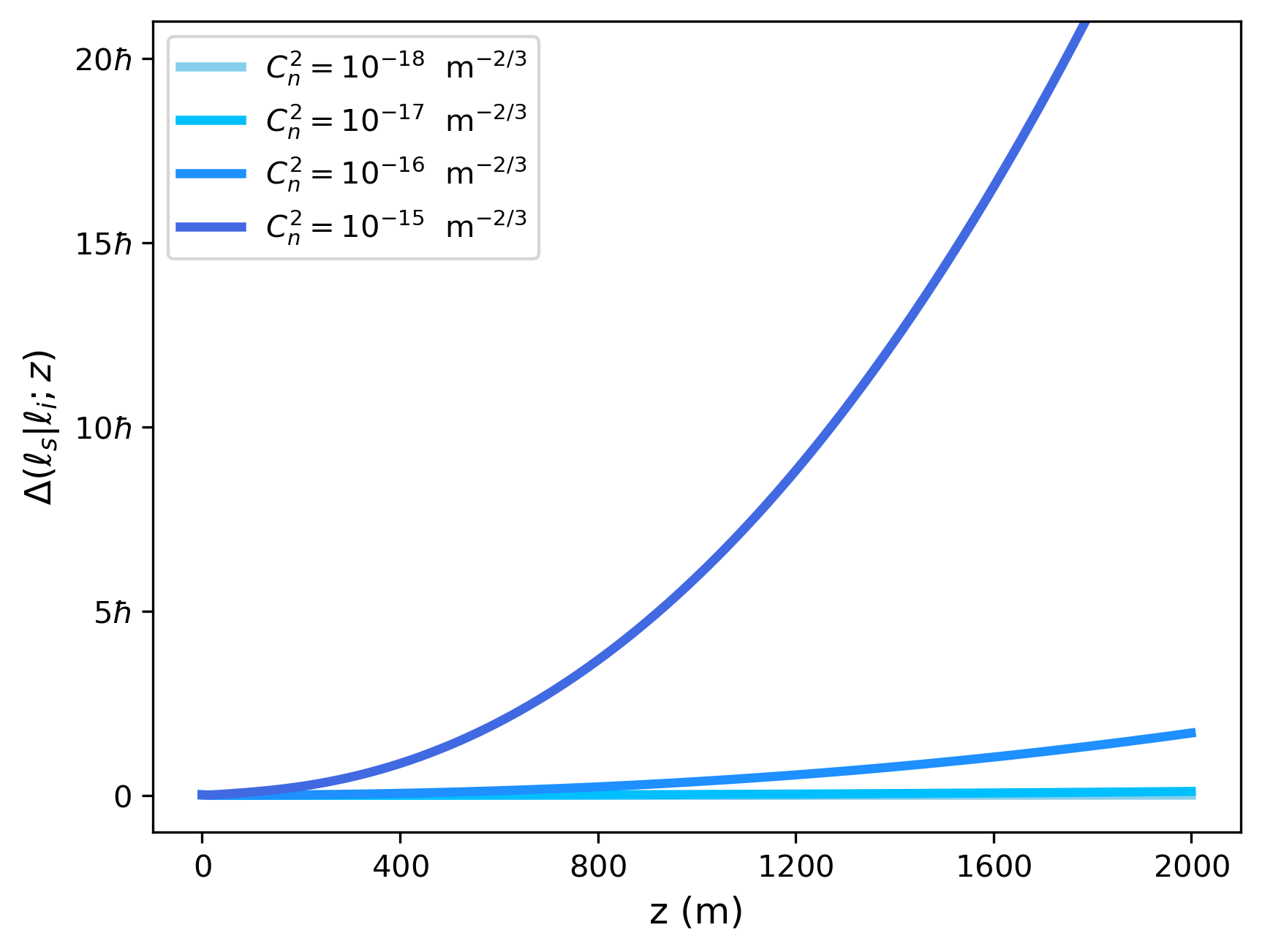}
    \caption{The standard deviation of the conditional OAM distribution $P(\ell_s|\ell_i=0;z)$ increases more rapidly with propagation distance z under stronger turbulence. However, under weak turbulence strengths ($C_n^2 = 10^{-17} - 10^{-16} \text{ m}^{-2/3}$), the standard deviation increases only linearly with z and remains small over a long distance. The parameter values are the same as those used in figure \ref{fig:oam_spectrum_z}. Note that the OAM is expressed in units of $\hbar$.}
    \label{fig:oam_width}
\end{figure}

The width of the conditional OAM distribution can be represented by its standard deviation, also known as the conditional OAM uncertainty. Figure \ref{fig:oam_width} illustrates the behavior of the conditional OAM uncertainty as a function of propagation distance $z$ under different turbulence strengths. Clearly, the conditional OAM uncertainty increases more rapidly under stronger turbulence. Furthermore, when the strength of the turbulence lies within a weak turbulence regime ($C_n^2 = 10^{-18} - 10^{-16} \text{ m}^{-2/3}$ in this case), the conditional OAM uncertainty remains relatively small over distances exceeding one kilometer. This analysis provides a useful benchmark for determining the practical distance limitations of free-space quantum communication links through the turbulent atmosphere.

\subsection{Finite distance range of maximal angle-OAM entanglement in turbulent channels}
\begin{figure}[htp]
    \centering
    \includegraphics[width=1\linewidth]{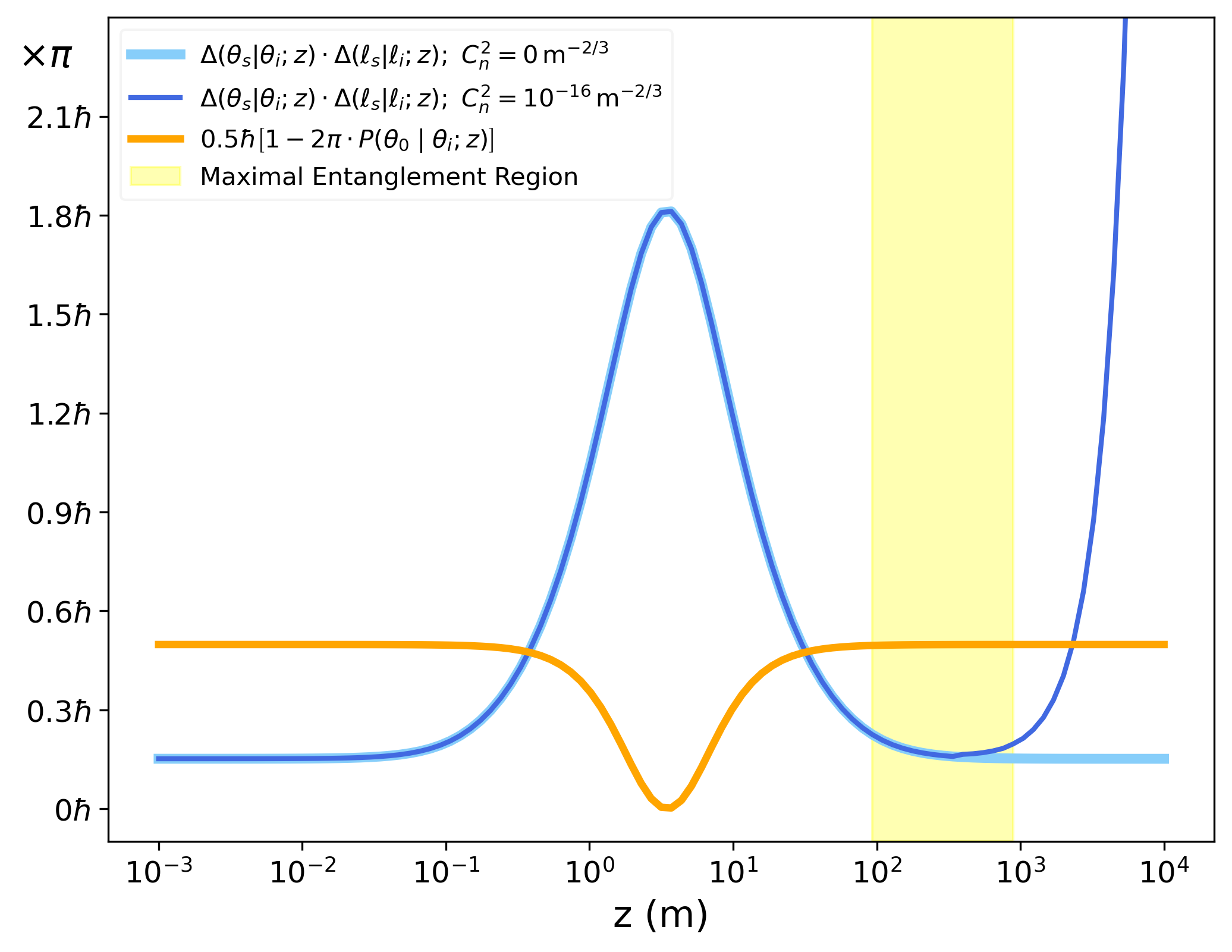}
    \caption{Conditional angle-OAM uncertainty product $\Delta (\theta_s | \theta_i=0;z) \cdot \Delta (\ell_s | \ell_i=0;z)$, and the right-hand side of the EPR criterion in the angle-OAM bases $0.5 \hbar \left[ 1 - 2\pi \cdot P(\theta_0 \mid \theta_i=0; z) \right]$ as a function of the propagation distance $z$. Entanglement in the angle-OAM bases exists when the blue curve falls below the orange curve. The range of distances indicated with yellow shading can be identified as the region of maximal entanglement. The parameter values are the same as those used in figure \ref{fig:p_cor}, and the turbulence strength is $C_n^2=10^{-16}\,\text{ m}^{-2/3}$. The conditional angle-OAM uncertainty product in the absence of turbulence is also shown, though note that the right-hand side of the EPR criterion is practically identical in both cases.}
    \label{fig:angle_oam_epr}
\end{figure}

Now we analyze the angle-OAM entanglement as the field propagates through turbulence using the EPR criterion in equation (\ref{equ:epr_angle}). The quantities $\Delta(\theta_s \mid \theta_i=0; z) \Delta(\ell_s \mid \ell_i=0; z)$ and $0.5 \hbar \left[ 1 - 2\pi P(\theta_0 \mid \theta_i=0; z) \right]$ are plotted together for comparison in figure \ref{fig:angle_oam_epr}. An additional term of $1\hbar$ is included in the conditional OAM uncertainty $\Delta(\ell_s \mid \ell_i=0; z)$ to account for practical imperfections \cite{bhattacharjee2022propagation}. At propagation distances where the blue curve lies below the orange curve, the signal and idler fields exhibit angle-OAM entanglement. Figure \ref{fig:angle_oam_epr} shows that the signal and idler fields are initially entangled, but that this entanglement disappears within a meter of propagation from the Fourier plane of the crystal. However, the angle-OAM entanglement revives at around $z=20 \text{ m}$ and persists over propagation distances of more than one kilometer. In the absence of turbulence, the angle-OAM entanglement would continue to persist after its revival regardless of how far the field continues to propagate, as shown in figure \ref{fig:angle_oam_epr}. These observations are consistent with the experimental results presented in a recent study of the propagation-induced revival of two-photon entanglement in the angle-OAM bases \cite{bhattacharjee2022propagation}, which explained the disappearance and revival (in the absence of turbulence) as a consequence of entanglement redistribution across different bases. Our analysis additionally shows that this entanglement can reappear and persist after long propagation distances through realistically-modeled turbulence, which has significant practical implications for free-space quantum communication. As was observed in \cite{bhattacharjee2022propagation}, our model also predicts a propagation distance at which the angle-OAM entanglement is maximized, which occurs when the EPR criterion in equation (\ref{equ:epr_angle}) is most strongly violated. Around this position, there exists a region (highlighted in yellow in figure \ref{fig:angle_oam_epr}) where the degree of entanglement is substantial enough to serve as a resource for quantum communication protocols operating over several kilometers through the weakly turbulent atmosphere.

\section{Discussion and Conclusion}
\label{sec:discussion}
We present a comprehensive theoretical study of the effects of weak turbulence on the propagation of photon pairs generated by spontaneous parametric down-conversion (SPDC). By employing the eHF principle together with the Kolmogorov model of turbulence, we develop a model to describe the propagation of the entire transverse profile of entangled photon pairs through homogeneous and isotropic turbulence, explicitly incorporating the evolution of the whole beam structure and the redistribution of entanglement among different degrees of freedom. Our model shows that turbulence can reduce the purity of the two-photon field but rather than eliminating transverse signal-idler spatial correlations entirely, it merely causes them to transition in nature from quantum to classical. Furthermore, turbulence leads to the broadening of the conditional OAM distribution, thus increasing the conditional OAM uncertainty. However, under relatively weak turbulence, the conditional OAM uncertainty remains low over long propagation distances, indicating the potential for the use of angle-OAM entanglement in practical free-space quantum communication protocols. As for this angle-OAM entanglement, our analysis reveals the existence of a finite range of distances over which this entanglement is maximized, allowing for the design and optimization of FSO quantum communication links operating over several kilometers in the turbulent atmosphere. As an example, the theoretical approach provided in this study can serve as a straightforward method for optimizing the link distance, as well as the properties of the transmitter.

Future research could employ more realistic models of field propagation through tubuelence, including non-Kolmogorov (anisotropic) turbulence and the multiple phase-screen approach. The double Gaussian approximation for the SPDC two-photon field could also be removed to further increase the accuracy of the model. It is also possible to combine the present theory with the design of adaptive optics to investigate optimal strategies for mitigating the detrimental effects of turbulence. 

\begin{acknowledgments}
This work was supported by the US Office of Naval Research award N00014-19-1-2247 and MURI award N00014-20-1-2558,  by the US National Science Foundation under Award 2138174, and by Cisco Systems, Inc. In addition, RWB acknowledges support by Natural Sciences and Engineering Research Council of Canada under award RGPIN/2017-06880, and the Canada First Research Excellence Fund award on Transformative Quantum Technologies under Award \#072623.
\end{acknowledgments}

\bibliography{bibliography}

\begin{thebibliography}{35}%
\makeatletter
\providecommand \@ifxundefined [1]{%
 \@ifx{#1\undefined}
}%
\providecommand \@ifnum [1]{%
 \ifnum #1\expandafter \@firstoftwo
 \else \expandafter \@secondoftwo
 \fi
}%
\providecommand \@ifx [1]{%
 \ifx #1\expandafter \@firstoftwo
 \else \expandafter \@secondoftwo
 \fi
}%
\providecommand \natexlab [1]{#1}%
\providecommand \enquote  [1]{``#1''}%
\providecommand \bibnamefont  [1]{#1}%
\providecommand \bibfnamefont [1]{#1}%
\providecommand \citenamefont [1]{#1}%
\providecommand \href@noop [0]{\@secondoftwo}%
\providecommand \href [0]{\begingroup \@sanitize@url \@href}%
\providecommand \@href[1]{\@@startlink{#1}\@@href}%
\providecommand \@@href[1]{\endgroup#1\@@endlink}%
\providecommand \@sanitize@url [0]{\catcode `\\12\catcode `\$12\catcode `\&12\catcode `\#12\catcode `\^12\catcode `\_12\catcode `\%12\relax}%
\providecommand \@@startlink[1]{}%
\providecommand \@@endlink[0]{}%
\providecommand \url  [0]{\begingroup\@sanitize@url \@url }%
\providecommand \@url [1]{\endgroup\@href {#1}{\urlprefix }}%
\providecommand \urlprefix  [0]{URL }%
\providecommand \Eprint [0]{\href }%
\providecommand \doibase [0]{https://doi.org/}%
\providecommand \selectlanguage [0]{\@gobble}%
\providecommand \bibinfo  [0]{\@secondoftwo}%
\providecommand \bibfield  [0]{\@secondoftwo}%
\providecommand \translation [1]{[#1]}%
\providecommand \BibitemOpen [0]{}%
\providecommand \bibitemStop [0]{}%
\providecommand \bibitemNoStop [0]{.\EOS\space}%
\providecommand \EOS [0]{\spacefactor3000\relax}%
\providecommand \BibitemShut  [1]{\csname bibitem#1\endcsname}%
\let\auto@bib@innerbib\@empty
\bibitem [{\citenamefont {Tsujino}\ \emph {et~al.}(2011)\citenamefont {Tsujino}, \citenamefont {Fukuda}, \citenamefont {Fujii}, \citenamefont {Inoue}, \citenamefont {Fujiwara}, \citenamefont {Takeoka},\ and\ \citenamefont {Sasaki}}]{tsujino2011quantum}%
  \BibitemOpen
  \bibfield  {author} {\bibinfo {author} {\bibfnamefont {K.}~\bibnamefont {Tsujino}}, \bibinfo {author} {\bibfnamefont {D.}~\bibnamefont {Fukuda}}, \bibinfo {author} {\bibfnamefont {G.}~\bibnamefont {Fujii}}, \bibinfo {author} {\bibfnamefont {S.}~\bibnamefont {Inoue}}, \bibinfo {author} {\bibfnamefont {M.}~\bibnamefont {Fujiwara}}, \bibinfo {author} {\bibfnamefont {M.}~\bibnamefont {Takeoka}},\ and\ \bibinfo {author} {\bibfnamefont {M.}~\bibnamefont {Sasaki}},\ }\bibfield  {title} {\bibinfo {title} {Quantum receiver beyond the standard quantum limit of coherent optical communication},\ }\href@noop {} {\bibfield  {journal} {\bibinfo  {journal} {Physical review letters}\ }\textbf {\bibinfo {volume} {106}},\ \bibinfo {pages} {250503} (\bibinfo {year} {2011})}\BibitemShut {NoStop}%
\bibitem [{\citenamefont {Jiang}\ \emph {et~al.}(2007)\citenamefont {Jiang}, \citenamefont {Taylor},\ and\ \citenamefont {Lukin}}]{jiang2007fast}%
  \BibitemOpen
  \bibfield  {author} {\bibinfo {author} {\bibfnamefont {L.}~\bibnamefont {Jiang}}, \bibinfo {author} {\bibfnamefont {J.}~\bibnamefont {Taylor}},\ and\ \bibinfo {author} {\bibfnamefont {M.}~\bibnamefont {Lukin}},\ }\bibfield  {title} {\bibinfo {title} {Fast and robust approach to long-distance quantum communication with atomic ensembles},\ }\href@noop {} {\bibfield  {journal} {\bibinfo  {journal} {Physical Review A}\ }\textbf {\bibinfo {volume} {76}},\ \bibinfo {pages} {012301} (\bibinfo {year} {2007})}\BibitemShut {NoStop}%
\bibitem [{\citenamefont {Yang}\ \emph {et~al.}(2021)\citenamefont {Yang}, \citenamefont {Liu}, \citenamefont {Gao}, \citenamefont {Zhou}, \citenamefont {Shi}, \citenamefont {Li},\ and\ \citenamefont {Li}}]{yang2021towards}%
  \BibitemOpen
  \bibfield  {author} {\bibinfo {author} {\bibfnamefont {Y.-G.}\ \bibnamefont {Yang}}, \bibinfo {author} {\bibfnamefont {X.-X.}\ \bibnamefont {Liu}}, \bibinfo {author} {\bibfnamefont {S.}~\bibnamefont {Gao}}, \bibinfo {author} {\bibfnamefont {Y.-H.}\ \bibnamefont {Zhou}}, \bibinfo {author} {\bibfnamefont {W.-M.}\ \bibnamefont {Shi}}, \bibinfo {author} {\bibfnamefont {J.}~\bibnamefont {Li}},\ and\ \bibinfo {author} {\bibfnamefont {D.}~\bibnamefont {Li}},\ }\bibfield  {title} {\bibinfo {title} {Towards practical anonymous quantum communication: {A} measurement-device-independent approach},\ }\href@noop {} {\bibfield  {journal} {\bibinfo  {journal} {Physical Review A}\ }\textbf {\bibinfo {volume} {104}},\ \bibinfo {pages} {052415} (\bibinfo {year} {2021})}\BibitemShut {NoStop}%
\bibitem [{\citenamefont {Mart{\'\i}nez}\ \emph {et~al.}(2018)\citenamefont {Mart{\'\i}nez}, \citenamefont {Tavakoli}, \citenamefont {Casanova}, \citenamefont {Canas}, \citenamefont {Marques},\ and\ \citenamefont {Lima}}]{martinez2018high}%
  \BibitemOpen
  \bibfield  {author} {\bibinfo {author} {\bibfnamefont {D.}~\bibnamefont {Mart{\'\i}nez}}, \bibinfo {author} {\bibfnamefont {A.}~\bibnamefont {Tavakoli}}, \bibinfo {author} {\bibfnamefont {M.}~\bibnamefont {Casanova}}, \bibinfo {author} {\bibfnamefont {G.}~\bibnamefont {Canas}}, \bibinfo {author} {\bibfnamefont {B.}~\bibnamefont {Marques}},\ and\ \bibinfo {author} {\bibfnamefont {G.}~\bibnamefont {Lima}},\ }\bibfield  {title} {\bibinfo {title} {High-dimensional quantum communication complexity beyond strategies based on {B}ell’s theorem},\ }\href@noop {} {\bibfield  {journal} {\bibinfo  {journal} {Physical review letters}\ }\textbf {\bibinfo {volume} {121}},\ \bibinfo {pages} {150504} (\bibinfo {year} {2018})}\BibitemShut {NoStop}%
\bibitem [{\citenamefont {Cao}\ \emph {et~al.}(2020)\citenamefont {Cao}, \citenamefont {Li}, \citenamefont {Yang}, \citenamefont {Jiang}, \citenamefont {Li}, \citenamefont {Hu}, \citenamefont {Abulizi}, \citenamefont {Li}, \citenamefont {Zhang}, \citenamefont {Sun} \emph {et~al.}}]{cao2020long}%
  \BibitemOpen
  \bibfield  {author} {\bibinfo {author} {\bibfnamefont {Y.}~\bibnamefont {Cao}}, \bibinfo {author} {\bibfnamefont {Y.-H.}\ \bibnamefont {Li}}, \bibinfo {author} {\bibfnamefont {K.-X.}\ \bibnamefont {Yang}}, \bibinfo {author} {\bibfnamefont {Y.-F.}\ \bibnamefont {Jiang}}, \bibinfo {author} {\bibfnamefont {S.-L.}\ \bibnamefont {Li}}, \bibinfo {author} {\bibfnamefont {X.-L.}\ \bibnamefont {Hu}}, \bibinfo {author} {\bibfnamefont {M.}~\bibnamefont {Abulizi}}, \bibinfo {author} {\bibfnamefont {C.-L.}\ \bibnamefont {Li}}, \bibinfo {author} {\bibfnamefont {W.}~\bibnamefont {Zhang}}, \bibinfo {author} {\bibfnamefont {Q.-C.}\ \bibnamefont {Sun}}, \emph {et~al.},\ }\bibfield  {title} {\bibinfo {title} {Long-distance free-space measurement-device-independent quantum key distribution},\ }\href@noop {} {\bibfield  {journal} {\bibinfo  {journal} {Physical Review Letters}\ }\textbf {\bibinfo {volume} {125}},\ \bibinfo {pages} {260503} (\bibinfo {year} {2020})}\BibitemShut {NoStop}%
\bibitem [{\citenamefont {Yang}\ \emph {et~al.}(2024)\citenamefont {Yang}, \citenamefont {Jiang}, \citenamefont {Benthin}, \citenamefont {Hanel}, \citenamefont {Fandrich}, \citenamefont {Joos}, \citenamefont {Bauer}, \citenamefont {Kolatschek}, \citenamefont {Hreibi}, \citenamefont {Rugeramigabo} \emph {et~al.}}]{yang2024high}%
  \BibitemOpen
  \bibfield  {author} {\bibinfo {author} {\bibfnamefont {J.}~\bibnamefont {Yang}}, \bibinfo {author} {\bibfnamefont {Z.}~\bibnamefont {Jiang}}, \bibinfo {author} {\bibfnamefont {F.}~\bibnamefont {Benthin}}, \bibinfo {author} {\bibfnamefont {J.}~\bibnamefont {Hanel}}, \bibinfo {author} {\bibfnamefont {T.}~\bibnamefont {Fandrich}}, \bibinfo {author} {\bibfnamefont {R.}~\bibnamefont {Joos}}, \bibinfo {author} {\bibfnamefont {S.}~\bibnamefont {Bauer}}, \bibinfo {author} {\bibfnamefont {S.}~\bibnamefont {Kolatschek}}, \bibinfo {author} {\bibfnamefont {A.}~\bibnamefont {Hreibi}}, \bibinfo {author} {\bibfnamefont {E.~P.}\ \bibnamefont {Rugeramigabo}}, \emph {et~al.},\ }\bibfield  {title} {\bibinfo {title} {High-rate intercity quantum key distribution with a semiconductor single-photon source},\ }\href@noop {} {\bibfield  {journal} {\bibinfo  {journal} {Light: Science \& Applications}\ }\textbf {\bibinfo {volume} {13}},\ \bibinfo {pages} {150} (\bibinfo {year} {2024})}\BibitemShut {NoStop}%
\bibitem [{\citenamefont {Chen}\ \emph {et~al.}(2021)\citenamefont {Chen}, \citenamefont {Zhang}, \citenamefont {Liu}, \citenamefont {Jiang}, \citenamefont {Zhang}, \citenamefont {Han}, \citenamefont {Ma}, \citenamefont {Hu}, \citenamefont {Li}, \citenamefont {Liu} \emph {et~al.}}]{chen2021twin}%
  \BibitemOpen
  \bibfield  {author} {\bibinfo {author} {\bibfnamefont {J.-P.}\ \bibnamefont {Chen}}, \bibinfo {author} {\bibfnamefont {C.}~\bibnamefont {Zhang}}, \bibinfo {author} {\bibfnamefont {Y.}~\bibnamefont {Liu}}, \bibinfo {author} {\bibfnamefont {C.}~\bibnamefont {Jiang}}, \bibinfo {author} {\bibfnamefont {W.-J.}\ \bibnamefont {Zhang}}, \bibinfo {author} {\bibfnamefont {Z.-Y.}\ \bibnamefont {Han}}, \bibinfo {author} {\bibfnamefont {S.-Z.}\ \bibnamefont {Ma}}, \bibinfo {author} {\bibfnamefont {X.-L.}\ \bibnamefont {Hu}}, \bibinfo {author} {\bibfnamefont {Y.-H.}\ \bibnamefont {Li}}, \bibinfo {author} {\bibfnamefont {H.}~\bibnamefont {Liu}}, \emph {et~al.},\ }\bibfield  {title} {\bibinfo {title} {Twin-field quantum key distribution over a 511 km optical fibre linking two distant metropolitan areas},\ }\href@noop {} {\bibfield  {journal} {\bibinfo  {journal} {Nature Photonics}\ }\textbf {\bibinfo {volume} {15}},\ \bibinfo {pages} {570} (\bibinfo {year} {2021})}\BibitemShut {NoStop}%
\bibitem [{\citenamefont {Cozzolino}\ \emph {et~al.}(2019)\citenamefont {Cozzolino}, \citenamefont {Bacco}, \citenamefont {Da~Lio}, \citenamefont {Ingerslev}, \citenamefont {Ding}, \citenamefont {Dalgaard}, \citenamefont {Kristensen}, \citenamefont {Galili}, \citenamefont {Rottwitt}, \citenamefont {Ramachandran} \emph {et~al.}}]{cozzolino2019orbital}%
  \BibitemOpen
  \bibfield  {author} {\bibinfo {author} {\bibfnamefont {D.}~\bibnamefont {Cozzolino}}, \bibinfo {author} {\bibfnamefont {D.}~\bibnamefont {Bacco}}, \bibinfo {author} {\bibfnamefont {B.}~\bibnamefont {Da~Lio}}, \bibinfo {author} {\bibfnamefont {K.}~\bibnamefont {Ingerslev}}, \bibinfo {author} {\bibfnamefont {Y.}~\bibnamefont {Ding}}, \bibinfo {author} {\bibfnamefont {K.}~\bibnamefont {Dalgaard}}, \bibinfo {author} {\bibfnamefont {P.}~\bibnamefont {Kristensen}}, \bibinfo {author} {\bibfnamefont {M.}~\bibnamefont {Galili}}, \bibinfo {author} {\bibfnamefont {K.}~\bibnamefont {Rottwitt}}, \bibinfo {author} {\bibfnamefont {S.}~\bibnamefont {Ramachandran}}, \emph {et~al.},\ }\bibfield  {title} {\bibinfo {title} {Orbital angular momentum states enabling fiber-based high-dimensional quantum communication},\ }\href@noop {} {\bibfield  {journal} {\bibinfo  {journal} {Physical Review Applied}\ }\textbf {\bibinfo {volume} {11}},\ \bibinfo {pages} {064058} (\bibinfo {year} {2019})}\BibitemShut {NoStop}%
\bibitem [{\citenamefont {Tan}\ \emph {et~al.}(2024)\citenamefont {Tan}, \citenamefont {Wang}, \citenamefont {Wu},\ and\ \citenamefont {He}}]{tan2024real}%
  \BibitemOpen
  \bibfield  {author} {\bibinfo {author} {\bibfnamefont {Y.}~\bibnamefont {Tan}}, \bibinfo {author} {\bibfnamefont {J.}~\bibnamefont {Wang}}, \bibinfo {author} {\bibfnamefont {J.}~\bibnamefont {Wu}},\ and\ \bibinfo {author} {\bibfnamefont {Z.}~\bibnamefont {He}},\ }\bibfield  {title} {\bibinfo {title} {Real-time polarization compensation method in quantum communication based on channel {M}uller parameters detection},\ }\href@noop {} {\bibfield  {journal} {\bibinfo  {journal} {Communications Engineering}\ }\textbf {\bibinfo {volume} {3}},\ \bibinfo {pages} {57} (\bibinfo {year} {2024})}\BibitemShut {NoStop}%
\bibitem [{\citenamefont {Dahal}\ and\ \citenamefont {Terno}(2021)}]{dahal2021polarization}%
  \BibitemOpen
  \bibfield  {author} {\bibinfo {author} {\bibfnamefont {P.~K.}\ \bibnamefont {Dahal}}\ and\ \bibinfo {author} {\bibfnamefont {D.~R.}\ \bibnamefont {Terno}},\ }\bibfield  {title} {\bibinfo {title} {Polarization rotation and near-{E}arth quantum communications},\ }\href@noop {} {\bibfield  {journal} {\bibinfo  {journal} {Physical Review A}\ }\textbf {\bibinfo {volume} {104}},\ \bibinfo {pages} {042610} (\bibinfo {year} {2021})}\BibitemShut {NoStop}%
\bibitem [{\citenamefont {Mafu}\ \emph {et~al.}(2013)\citenamefont {Mafu}, \citenamefont {Dudley}, \citenamefont {Goyal}, \citenamefont {Giovannini}, \citenamefont {McLaren}, \citenamefont {Padgett}, \citenamefont {Konrad}, \citenamefont {Petruccione}, \citenamefont {L{\"u}tkenhaus},\ and\ \citenamefont {Forbes}}]{mafu2013higher}%
  \BibitemOpen
  \bibfield  {author} {\bibinfo {author} {\bibfnamefont {M.}~\bibnamefont {Mafu}}, \bibinfo {author} {\bibfnamefont {A.}~\bibnamefont {Dudley}}, \bibinfo {author} {\bibfnamefont {S.}~\bibnamefont {Goyal}}, \bibinfo {author} {\bibfnamefont {D.}~\bibnamefont {Giovannini}}, \bibinfo {author} {\bibfnamefont {M.}~\bibnamefont {McLaren}}, \bibinfo {author} {\bibfnamefont {M.~J.}\ \bibnamefont {Padgett}}, \bibinfo {author} {\bibfnamefont {T.}~\bibnamefont {Konrad}}, \bibinfo {author} {\bibfnamefont {F.}~\bibnamefont {Petruccione}}, \bibinfo {author} {\bibfnamefont {N.}~\bibnamefont {L{\"u}tkenhaus}},\ and\ \bibinfo {author} {\bibfnamefont {A.}~\bibnamefont {Forbes}},\ }\bibfield  {title} {\bibinfo {title} {Higher-dimensional orbital-angular-momentum-based quantum key distribution with mutually unbiased bases},\ }\href@noop {} {\bibfield  {journal} {\bibinfo  {journal} {Physical Review A}\ }\textbf {\bibinfo {volume} {88}},\ \bibinfo {pages} {032305} (\bibinfo {year} {2013})}\BibitemShut {NoStop}%
\bibitem [{\citenamefont {Xu}\ \emph {et~al.}(2024)\citenamefont {Xu}, \citenamefont {Choudhary},\ and\ \citenamefont {Boyd}}]{xu2024stimulated}%
  \BibitemOpen
  \bibfield  {author} {\bibinfo {author} {\bibfnamefont {Y.}~\bibnamefont {Xu}}, \bibinfo {author} {\bibfnamefont {S.}~\bibnamefont {Choudhary}},\ and\ \bibinfo {author} {\bibfnamefont {R.~W.}\ \bibnamefont {Boyd}},\ }\bibfield  {title} {\bibinfo {title} {Stimulated emission tomography for efficient characterization of spatial entanglement},\ }\href@noop {} {\bibfield  {journal} {\bibinfo  {journal} {Physical Review Research}\ }\textbf {\bibinfo {volume} {6}},\ \bibinfo {pages} {L042047} (\bibinfo {year} {2024})}\BibitemShut {NoStop}%
\bibitem [{\citenamefont {Lo}\ \emph {et~al.}(2014)\citenamefont {Lo}, \citenamefont {Curty},\ and\ \citenamefont {Tamaki}}]{lo2014secure}%
  \BibitemOpen
  \bibfield  {author} {\bibinfo {author} {\bibfnamefont {H.-K.}\ \bibnamefont {Lo}}, \bibinfo {author} {\bibfnamefont {M.}~\bibnamefont {Curty}},\ and\ \bibinfo {author} {\bibfnamefont {K.}~\bibnamefont {Tamaki}},\ }\bibfield  {title} {\bibinfo {title} {Secure quantum key distribution},\ }\href@noop {} {\bibfield  {journal} {\bibinfo  {journal} {Nature Photonics}\ }\textbf {\bibinfo {volume} {8}},\ \bibinfo {pages} {595} (\bibinfo {year} {2014})}\BibitemShut {NoStop}%
\bibitem [{\citenamefont {Tyler}\ and\ \citenamefont {Boyd}(2009)}]{tyler2009influence}%
  \BibitemOpen
  \bibfield  {author} {\bibinfo {author} {\bibfnamefont {G.~A.}\ \bibnamefont {Tyler}}\ and\ \bibinfo {author} {\bibfnamefont {R.~W.}\ \bibnamefont {Boyd}},\ }\bibfield  {title} {\bibinfo {title} {Influence of atmospheric turbulence on the propagation of quantum states of light carrying orbital angular momentum},\ }\href@noop {} {\bibfield  {journal} {\bibinfo  {journal} {Optics letters}\ }\textbf {\bibinfo {volume} {34}},\ \bibinfo {pages} {142} (\bibinfo {year} {2009})}\BibitemShut {NoStop}%
\bibitem [{\citenamefont {Yan}\ \emph {et~al.}(2016)\citenamefont {Yan}, \citenamefont {Zhang}, \citenamefont {Zhang}, \citenamefont {Chun},\ and\ \citenamefont {Fan}}]{yan2016decoherence}%
  \BibitemOpen
  \bibfield  {author} {\bibinfo {author} {\bibfnamefont {X.}~\bibnamefont {Yan}}, \bibinfo {author} {\bibfnamefont {P.~F.}\ \bibnamefont {Zhang}}, \bibinfo {author} {\bibfnamefont {J.~H.}\ \bibnamefont {Zhang}}, \bibinfo {author} {\bibfnamefont {H.~Q.}\ \bibnamefont {Chun}},\ and\ \bibinfo {author} {\bibfnamefont {C.~Y.}\ \bibnamefont {Fan}},\ }\bibfield  {title} {\bibinfo {title} {Decoherence of orbital angular momentum tangled photons in non-{K}olmogorov turbulence},\ }\href@noop {} {\bibfield  {journal} {\bibinfo  {journal} {JOSA A}\ }\textbf {\bibinfo {volume} {33}},\ \bibinfo {pages} {1831} (\bibinfo {year} {2016})}\BibitemShut {NoStop}%
\bibitem [{\citenamefont {Jha}\ \emph {et~al.}(2010)\citenamefont {Jha}, \citenamefont {Tyler},\ and\ \citenamefont {Boyd}}]{jha2010effects}%
  \BibitemOpen
  \bibfield  {author} {\bibinfo {author} {\bibfnamefont {A.~K.}\ \bibnamefont {Jha}}, \bibinfo {author} {\bibfnamefont {G.~A.}\ \bibnamefont {Tyler}},\ and\ \bibinfo {author} {\bibfnamefont {R.~W.}\ \bibnamefont {Boyd}},\ }\bibfield  {title} {\bibinfo {title} {Effects of atmospheric turbulence on the entanglement of spatial two-qubit states},\ }\href@noop {} {\bibfield  {journal} {\bibinfo  {journal} {Physical Review A}\ }\textbf {\bibinfo {volume} {81}},\ \bibinfo {pages} {053832} (\bibinfo {year} {2010})}\BibitemShut {NoStop}%
\bibitem [{\citenamefont {Roux}\ \emph {et~al.}(2015)\citenamefont {Roux}, \citenamefont {Wellens},\ and\ \citenamefont {Shatokhin}}]{roux2015entanglement}%
  \BibitemOpen
  \bibfield  {author} {\bibinfo {author} {\bibfnamefont {F.~S.}\ \bibnamefont {Roux}}, \bibinfo {author} {\bibfnamefont {T.}~\bibnamefont {Wellens}},\ and\ \bibinfo {author} {\bibfnamefont {V.~N.}\ \bibnamefont {Shatokhin}},\ }\bibfield  {title} {\bibinfo {title} {Entanglement evolution of twisted photons in strong atmospheric turbulence},\ }\href@noop {} {\bibfield  {journal} {\bibinfo  {journal} {Physical Review A}\ }\textbf {\bibinfo {volume} {92}},\ \bibinfo {pages} {012326} (\bibinfo {year} {2015})}\BibitemShut {NoStop}%
\bibitem [{\citenamefont {Ibrahim}\ \emph {et~al.}(2014)\citenamefont {Ibrahim}, \citenamefont {Roux},\ and\ \citenamefont {Konrad}}]{ibrahim2014parameter}%
  \BibitemOpen
  \bibfield  {author} {\bibinfo {author} {\bibfnamefont {A.~H.}\ \bibnamefont {Ibrahim}}, \bibinfo {author} {\bibfnamefont {F.~S.}\ \bibnamefont {Roux}},\ and\ \bibinfo {author} {\bibfnamefont {T.}~\bibnamefont {Konrad}},\ }\bibfield  {title} {\bibinfo {title} {Parameter dependence in the atmospheric decoherence of modally entangled photon pairs},\ }\href@noop {} {\bibfield  {journal} {\bibinfo  {journal} {Physical Review A}\ }\textbf {\bibinfo {volume} {90}},\ \bibinfo {pages} {052115} (\bibinfo {year} {2014})}\BibitemShut {NoStop}%
\bibitem [{\citenamefont {Chan}\ \emph {et~al.}(2007)\citenamefont {Chan}, \citenamefont {Torres},\ and\ \citenamefont {Eberly}}]{chan2007transverse}%
  \BibitemOpen
  \bibfield  {author} {\bibinfo {author} {\bibfnamefont {K.}~\bibnamefont {Chan}}, \bibinfo {author} {\bibfnamefont {J.}~\bibnamefont {Torres}},\ and\ \bibinfo {author} {\bibfnamefont {J.~H.}\ \bibnamefont {Eberly}},\ }\bibfield  {title} {\bibinfo {title} {Transverse entanglement migration in {H}ilbert space},\ }\href@noop {} {\bibfield  {journal} {\bibinfo  {journal} {Physical Review A}\ }\textbf {\bibinfo {volume} {75}},\ \bibinfo {pages} {050101} (\bibinfo {year} {2007})}\BibitemShut {NoStop}%
\bibitem [{\citenamefont {Reichert}\ \emph {et~al.}(2017)\citenamefont {Reichert}, \citenamefont {Sun},\ and\ \citenamefont {Fleischer}}]{reichert2017quality}%
  \BibitemOpen
  \bibfield  {author} {\bibinfo {author} {\bibfnamefont {M.}~\bibnamefont {Reichert}}, \bibinfo {author} {\bibfnamefont {X.}~\bibnamefont {Sun}},\ and\ \bibinfo {author} {\bibfnamefont {J.~W.}\ \bibnamefont {Fleischer}},\ }\bibfield  {title} {\bibinfo {title} {Quality of spatial entanglement propagation},\ }\href@noop {} {\bibfield  {journal} {\bibinfo  {journal} {Physical Review A}\ }\textbf {\bibinfo {volume} {95}},\ \bibinfo {pages} {063836} (\bibinfo {year} {2017})}\BibitemShut {NoStop}%
\bibitem [{\citenamefont {Bhattacharjee}\ \emph {et~al.}(2022)\citenamefont {Bhattacharjee}, \citenamefont {Joshi}, \citenamefont {Karan}, \citenamefont {Leach},\ and\ \citenamefont {Jha}}]{bhattacharjee2022propagation}%
  \BibitemOpen
  \bibfield  {author} {\bibinfo {author} {\bibfnamefont {A.}~\bibnamefont {Bhattacharjee}}, \bibinfo {author} {\bibfnamefont {M.~K.}\ \bibnamefont {Joshi}}, \bibinfo {author} {\bibfnamefont {S.}~\bibnamefont {Karan}}, \bibinfo {author} {\bibfnamefont {J.}~\bibnamefont {Leach}},\ and\ \bibinfo {author} {\bibfnamefont {A.~K.}\ \bibnamefont {Jha}},\ }\bibfield  {title} {\bibinfo {title} {Propagation-induced revival of entanglement in the angle-{OAM} bases},\ }\href@noop {} {\bibfield  {journal} {\bibinfo  {journal} {Science Advances}\ }\textbf {\bibinfo {volume} {8}},\ \bibinfo {pages} {eabn7876} (\bibinfo {year} {2022})}\BibitemShut {NoStop}%
\bibitem [{\citenamefont {Just}\ \emph {et~al.}(2013)\citenamefont {Just}, \citenamefont {Cavanna}, \citenamefont {Chekhova},\ and\ \citenamefont {Leuchs}}]{just2013transverse}%
  \BibitemOpen
  \bibfield  {author} {\bibinfo {author} {\bibfnamefont {F.}~\bibnamefont {Just}}, \bibinfo {author} {\bibfnamefont {A.}~\bibnamefont {Cavanna}}, \bibinfo {author} {\bibfnamefont {M.~V.}\ \bibnamefont {Chekhova}},\ and\ \bibinfo {author} {\bibfnamefont {G.}~\bibnamefont {Leuchs}},\ }\bibfield  {title} {\bibinfo {title} {Transverse entanglement of biphotons},\ }\href@noop {} {\bibfield  {journal} {\bibinfo  {journal} {New Journal of Physics}\ }\textbf {\bibinfo {volume} {15}},\ \bibinfo {pages} {083015} (\bibinfo {year} {2013})}\BibitemShut {NoStop}%
\bibitem [{\citenamefont {Wang}\ and\ \citenamefont {Plonus}(1979)}]{wang1979optical}%
  \BibitemOpen
  \bibfield  {author} {\bibinfo {author} {\bibfnamefont {S.}~\bibnamefont {Wang}}\ and\ \bibinfo {author} {\bibfnamefont {M.}~\bibnamefont {Plonus}},\ }\bibfield  {title} {\bibinfo {title} {Optical beam propagation for a partially coherent source in the turbulent atmosphere},\ }\href@noop {} {\bibfield  {journal} {\bibinfo  {journal} {JOSA}\ }\textbf {\bibinfo {volume} {69}},\ \bibinfo {pages} {1297} (\bibinfo {year} {1979})}\BibitemShut {NoStop}%
\bibitem [{\citenamefont {Andrews}\ and\ \citenamefont {Phillips}(2005)}]{andrews2005laser}%
  \BibitemOpen
  \bibfield  {author} {\bibinfo {author} {\bibfnamefont {L.~C.}\ \bibnamefont {Andrews}}\ and\ \bibinfo {author} {\bibfnamefont {R.~L.}\ \bibnamefont {Phillips}},\ }\href@noop {} {\emph {\bibinfo {title} {Laser {B}eam {P}ropagation {T}hrough {R}andom {M}edia, {S}econd edition}}}\ (\bibinfo  {publisher} {SPIE Optical Engineering Press},\ \bibinfo {year} {2005})\BibitemShut {NoStop}%
\bibitem [{\citenamefont {Wang}\ \emph {et~al.}(2015)\citenamefont {Wang}, \citenamefont {Liu},\ and\ \citenamefont {Cai}}]{wang2015propagation}%
  \BibitemOpen
  \bibfield  {author} {\bibinfo {author} {\bibfnamefont {F.}~\bibnamefont {Wang}}, \bibinfo {author} {\bibfnamefont {X.}~\bibnamefont {Liu}},\ and\ \bibinfo {author} {\bibfnamefont {Y.}~\bibnamefont {Cai}},\ }\bibfield  {title} {\bibinfo {title} {Propagation of partially coherent beam in turbulent atmosphere: a review (invited review)},\ }\href@noop {} {\bibfield  {journal} {\bibinfo  {journal} {Progress In Electromagnetics Research}\ }\textbf {\bibinfo {volume} {150}},\ \bibinfo {pages} {123} (\bibinfo {year} {2015})}\BibitemShut {NoStop}%
\bibitem [{\citenamefont {Bochove}\ and\ \citenamefont {Rao~Gudimetla}(2016)}]{bochove2016approach}%
  \BibitemOpen
  \bibfield  {author} {\bibinfo {author} {\bibfnamefont {E.~J.}\ \bibnamefont {Bochove}}\ and\ \bibinfo {author} {\bibfnamefont {V.}~\bibnamefont {Rao~Gudimetla}},\ }\bibfield  {title} {\bibinfo {title} {Approach to atmospheric laser-propagation theory based on the extended huygens--{F}resnel principle and a self-consistency concept},\ }\href@noop {} {\bibfield  {journal} {\bibinfo  {journal} {Journal of the Optical Society of America A}\ }\textbf {\bibinfo {volume} {34}},\ \bibinfo {pages} {140} (\bibinfo {year} {2016})}\BibitemShut {NoStop}%
\bibitem [{\citenamefont {Ponomarenko}(2022)}]{ponomarenko2022classical}%
  \BibitemOpen
  \bibfield  {author} {\bibinfo {author} {\bibfnamefont {S.~A.}\ \bibnamefont {Ponomarenko}},\ }\bibfield  {title} {\bibinfo {title} {Classical entanglement of twisted random light propagating through atmospheric turbulence},\ }\href@noop {} {\bibfield  {journal} {\bibinfo  {journal} {JOSA A}\ }\textbf {\bibinfo {volume} {39}},\ \bibinfo {pages} {C1} (\bibinfo {year} {2022})}\BibitemShut {NoStop}%
\bibitem [{\citenamefont {Yura}(1972)}]{yura1972mutual}%
  \BibitemOpen
  \bibfield  {author} {\bibinfo {author} {\bibfnamefont {H.}~\bibnamefont {Yura}},\ }\bibfield  {title} {\bibinfo {title} {Mutual coherence function of a finite cross section optical beam propagating in a turbulent medium},\ }\href@noop {} {\bibfield  {journal} {\bibinfo  {journal} {Applied Optics}\ }\textbf {\bibinfo {volume} {11}},\ \bibinfo {pages} {1399} (\bibinfo {year} {1972})}\BibitemShut {NoStop}%
\bibitem [{\citenamefont {Phehlukwayo}\ \emph {et~al.}(2020)\citenamefont {Phehlukwayo}, \citenamefont {Umuhire}, \citenamefont {Ismail}, \citenamefont {Joshi},\ and\ \citenamefont {Petruccione}}]{phehlukwayo2020influence}%
  \BibitemOpen
  \bibfield  {author} {\bibinfo {author} {\bibfnamefont {S.~P.}\ \bibnamefont {Phehlukwayo}}, \bibinfo {author} {\bibfnamefont {M.~L.}\ \bibnamefont {Umuhire}}, \bibinfo {author} {\bibfnamefont {Y.}~\bibnamefont {Ismail}}, \bibinfo {author} {\bibfnamefont {S.}~\bibnamefont {Joshi}},\ and\ \bibinfo {author} {\bibfnamefont {F.}~\bibnamefont {Petruccione}},\ }\bibfield  {title} {\bibinfo {title} {Influence of coincidence detection of a biphoton state through free-space atmospheric turbulence using a partially spatially coherent pump},\ }\href@noop {} {\bibfield  {journal} {\bibinfo  {journal} {Physical Review A}\ }\textbf {\bibinfo {volume} {102}},\ \bibinfo {pages} {033732} (\bibinfo {year} {2020})}\BibitemShut {NoStop}%
\bibitem [{\citenamefont {Schneeloch}\ and\ \citenamefont {Howell}(2016)}]{schneeloch2016introduction}%
  \BibitemOpen
  \bibfield  {author} {\bibinfo {author} {\bibfnamefont {J.}~\bibnamefont {Schneeloch}}\ and\ \bibinfo {author} {\bibfnamefont {J.~C.}\ \bibnamefont {Howell}},\ }\bibfield  {title} {\bibinfo {title} {Introduction to the transverse spatial correlations in spontaneous parametric down-conversion through the biphoton birth zone},\ }\href@noop {} {\bibfield  {journal} {\bibinfo  {journal} {Journal of Optics}\ }\textbf {\bibinfo {volume} {18}},\ \bibinfo {pages} {053501} (\bibinfo {year} {2016})}\BibitemShut {NoStop}%
\bibitem [{\citenamefont {Walborn}\ \emph {et~al.}(2010)\citenamefont {Walborn}, \citenamefont {Monken}, \citenamefont {P{\'a}dua},\ and\ \citenamefont {Ribeiro}}]{walborn2010spatial}%
  \BibitemOpen
  \bibfield  {author} {\bibinfo {author} {\bibfnamefont {S.~P.}\ \bibnamefont {Walborn}}, \bibinfo {author} {\bibfnamefont {C.}~\bibnamefont {Monken}}, \bibinfo {author} {\bibfnamefont {S.}~\bibnamefont {P{\'a}dua}},\ and\ \bibinfo {author} {\bibfnamefont {P.~S.}\ \bibnamefont {Ribeiro}},\ }\bibfield  {title} {\bibinfo {title} {Spatial correlations in parametric down-conversion},\ }\href@noop {} {\bibfield  {journal} {\bibinfo  {journal} {Physics Reports}\ }\textbf {\bibinfo {volume} {495}},\ \bibinfo {pages} {87} (\bibinfo {year} {2010})}\BibitemShut {NoStop}%
\bibitem [{\citenamefont {Edgar}\ \emph {et~al.}(2012)\citenamefont {Edgar}, \citenamefont {Tasca}, \citenamefont {Izdebski}, \citenamefont {Warburton}, \citenamefont {Leach}, \citenamefont {Agnew}, \citenamefont {Buller}, \citenamefont {Boyd},\ and\ \citenamefont {Padgett}}]{edgar2012imaging}%
  \BibitemOpen
  \bibfield  {author} {\bibinfo {author} {\bibfnamefont {M.~P.}\ \bibnamefont {Edgar}}, \bibinfo {author} {\bibfnamefont {D.~S.}\ \bibnamefont {Tasca}}, \bibinfo {author} {\bibfnamefont {F.}~\bibnamefont {Izdebski}}, \bibinfo {author} {\bibfnamefont {R.~E.}\ \bibnamefont {Warburton}}, \bibinfo {author} {\bibfnamefont {J.}~\bibnamefont {Leach}}, \bibinfo {author} {\bibfnamefont {M.}~\bibnamefont {Agnew}}, \bibinfo {author} {\bibfnamefont {G.~S.}\ \bibnamefont {Buller}}, \bibinfo {author} {\bibfnamefont {R.~W.}\ \bibnamefont {Boyd}},\ and\ \bibinfo {author} {\bibfnamefont {M.~J.}\ \bibnamefont {Padgett}},\ }\bibfield  {title} {\bibinfo {title} {Imaging high-dimensional spatial entanglement with a camera},\ }\href@noop {} {\bibfield  {journal} {\bibinfo  {journal} {Nature communications}\ }\textbf {\bibinfo {volume} {3}},\ \bibinfo {pages} {984} (\bibinfo {year} {2012})}\BibitemShut {NoStop}%
\bibitem [{\citenamefont {Wang}\ \emph {et~al.}(1983)\citenamefont {Wang}, \citenamefont {Baykal},\ and\ \citenamefont {Plonus}}]{wang1983receiver}%
  \BibitemOpen
  \bibfield  {author} {\bibinfo {author} {\bibfnamefont {S.}~\bibnamefont {Wang}}, \bibinfo {author} {\bibfnamefont {Y.}~\bibnamefont {Baykal}},\ and\ \bibinfo {author} {\bibfnamefont {M.}~\bibnamefont {Plonus}},\ }\bibfield  {title} {\bibinfo {title} {Receiver-aperture averaging effects for the intensity fluctuation of a beam wave in the turbulent atmosphere},\ }\href@noop {} {\bibfield  {journal} {\bibinfo  {journal} {Journal of the Optical Society of America}\ }\textbf {\bibinfo {volume} {73}},\ \bibinfo {pages} {831} (\bibinfo {year} {1983})}\BibitemShut {NoStop}%
\bibitem [{\citenamefont {Leach}\ \emph {et~al.}(2010)\citenamefont {Leach}, \citenamefont {Jack}, \citenamefont {Romero}, \citenamefont {Jha}, \citenamefont {Yao}, \citenamefont {Franke-Arnold}, \citenamefont {Ireland}, \citenamefont {Boyd}, \citenamefont {Barnett},\ and\ \citenamefont {Padgett}}]{leach2010quantum}%
  \BibitemOpen
  \bibfield  {author} {\bibinfo {author} {\bibfnamefont {J.}~\bibnamefont {Leach}}, \bibinfo {author} {\bibfnamefont {B.}~\bibnamefont {Jack}}, \bibinfo {author} {\bibfnamefont {J.}~\bibnamefont {Romero}}, \bibinfo {author} {\bibfnamefont {A.~K.}\ \bibnamefont {Jha}}, \bibinfo {author} {\bibfnamefont {A.~M.}\ \bibnamefont {Yao}}, \bibinfo {author} {\bibfnamefont {S.}~\bibnamefont {Franke-Arnold}}, \bibinfo {author} {\bibfnamefont {D.~G.}\ \bibnamefont {Ireland}}, \bibinfo {author} {\bibfnamefont {R.~W.}\ \bibnamefont {Boyd}}, \bibinfo {author} {\bibfnamefont {S.~M.}\ \bibnamefont {Barnett}},\ and\ \bibinfo {author} {\bibfnamefont {M.~J.}\ \bibnamefont {Padgett}},\ }\bibfield  {title} {\bibinfo {title} {Quantum correlations in optical angle--orbital angular momentum variables},\ }\href@noop {} {\bibfield  {journal} {\bibinfo  {journal} {Science}\ }\textbf {\bibinfo {volume} {329}},\ \bibinfo {pages} {662} (\bibinfo {year} {2010})}\BibitemShut {NoStop}%
\bibitem [{\citenamefont {Schneeloch}\ and\ \citenamefont {Howland}(2018)}]{schneeloch2018quantifying}%
  \BibitemOpen
  \bibfield  {author} {\bibinfo {author} {\bibfnamefont {J.}~\bibnamefont {Schneeloch}}\ and\ \bibinfo {author} {\bibfnamefont {G.~A.}\ \bibnamefont {Howland}},\ }\bibfield  {title} {\bibinfo {title} {Quantifying high-dimensional entanglement with {E}instein-{P}odolsky-{R}osen correlations},\ }\href@noop {} {\bibfield  {journal} {\bibinfo  {journal} {Physical Review A}\ }\textbf {\bibinfo {volume} {97}},\ \bibinfo {pages} {042338} (\bibinfo {year} {2018})}\BibitemShut {NoStop}%
\end{thebibliography}%

\end{document}